\documentclass[aps,11pt,letterpaper]{revtex4}
\usepackage{amsmath,amssymb,mathptmx,slashed,bm}
\usepackage[pdftex]{graphicx}
\newcommand{\abbrev}{\small}
\newcommand{\vp}[1]{\vphantom{#1}}
\newcommand{\Lx}{\left(}
\newcommand{\Rx}{\right)}
\newcommand{\LB}{\left[}
\newcommand{\RB}{\right]}
\newcommand{\hypgeo}[4]{{\vphantom{F}}_2F_1\Lx{#1},{\,#2};{\,#3};{\,#4}\Rx}
\newcommand{\ghypgeo}[6]{{\vphantom{F}}_3F_2\Lx{#1},{\,#2},{\,#3};{\,#4},{\,#5};{\,#6}\Rx}

\newcommand{\cg}{{c_\Gamma}}

\newcommand{\sos}[2]{\frac{s_{#1}}{s_{#2}}}
\newcommand{\ep}{{\varepsilon}}
\newcommand{\ve}{{\epsilon}}
\newcommand{\eqn}[1]{Eq.\,(\ref{#1})}
\newcommand{\eqns}[2]{Eqs.\,(\ref{#1}-\ref{#2})}
\newcommand{\fig}[1]{Fig.\,(\ref{#1})}

\newcommand{\Sec}[1]{Section~\ref{#1}}
\newcommand{\order}[1]{{\cal O}(#1)}

\newcommand{\Li}[1]{\mathop{\hbox{\rm Li}_{#1}}\nolimits}

\newcommand{\nlo}{{NLO}}
\newcommand{\nnlo}{{NNLO}}
\newcommand{\nnnlo}{{N${}^3$LO}}
\newcommand{\qcd} {{QCD}}

\newcommand{\bare}{{\rm B}}

\newcommand{\logmut}[1]{l_{t}^{#1}}

\newcommand{\ms} {{\rm MS}}

\newcommand{\msbar} {{\overline{\ms}}}

\newcommand{\as} {{\alpha_s}}
\newcommand{\aspi} {{\Lx\frac{\alpha_s}{\pi}\Rx}}
\newcommand{\asbare} {{\alpha_s^{B}}}

\newcommand{\amsbar} {{\alpha_{s}^{\msbar}}}

\newcommand{\amsbarpi} {{\Lx\frac{\alpha_{s}^{\msbar}}{\pi}\Rx}}

\newcommand{\betabar}[1] {{\beta_{#1}^{\msbar}}}

\newcommand{\xb}{\bar{x}\,}
\newcommand{\yb}{\bar{y}\,}

\newcommand{\REDUZE}{{\abbrev REDUZE2}}
\newcommand{\QGRAF}{{\abbrev QGRAF}}

\newcommand{\HypExp}{{\abbrev HypExp}}
\newcommand{\FORM}{{\abbrev FORM}}

\newfont{\scyr}{wncyr10 scaled 550}
\newcommand{\shuf}{\mbox{\bf \scyr X}}
\newcommand{\abare}{{\Lx\frac{g^2(4\,\pi)^{\ep}}{4\,\pi^2\,\Gamma(1-\ep)}\Rx}}
\newcommand{\abarex}{{\Lx\frac{g^2(4\,\pi)^{\ep}}{4\,\pi^2\,\exp(\ep\,\gamma_E)}\Rx}}
\newcommand{\DSum}[1] {{\cal D}^{#1}(\xb\!)}
\newcommand{\mumh} {{\Lx\frac{\mu^2}{M_H^2}\Rx}}
\newcommand{\Cfact} {{\frac{C_1^2\,\pi}{64\,v^2}}}

\begin{document}

\today

\bibliographystyle{apsrev}

\title{One-Loop Single-Real-Emission Contributions to $pp\to H + X$ at
  Next-to-Next-to-Next-to-Leading Order}

\author{William~B.~Kilgore}
\affiliation{Physics Department, Brookhaven National Laboratory,
  Upton, New York
  11973, USA.\\
  {\tt [kilgore@bnl.gov]} }

\begin{abstract}
  I compute the contributions of the one-loop single-real-emission
  amplitudes, $gg\to H g$, $qg\to H q$, etc., to inclusive Higgs boson
  production through next-to-next-to-next-to-leading order (\nnnlo) in
  the strong coupling $\as$.  The next-to-leading (\nlo) and
  next-to-next-to-leading order (\nnlo) terms are computed in closed
  form, in terms of $\Gamma$-functions and the hypergeometric
  functions ${}_{2}F_{1}$ and ${}_{3}F_{2}$.  I compute the \nnnlo\
  terms as Laurent expansions in the dimensional regularization
  parameter through order $(\ep^{1})$.  To obtain the \nnnlo\ terms, I
  perform an extended threshold expansion of the phase space integrals
  and map the resulting coefficients onto a basis of harmonic
  polylogarithms.
\end{abstract}

\maketitle

\section{Introduction}
On July 4, 2012, the ATLAS and CMS collaborations at the CERN Large
Hadron Collider (LHC) announced the discovery of a new particle with
mass near $126$ GeV~\cite{Aad:2012tfa,Chatrchyan:2012ufa}.  The
initial discovery and subsequent measurements indicate that this new
particle looks very much like the long-anticipated Higgs
boson~\cite{Aad:2013wqa,Aad:2013xqa,Chatrchyan:2012jja,Chatrchyan:2013lba}.
It is of the first importance to determine if this discovery is indeed
the Higgs boson of the Standard Model, a component of a more
complicated symmetry-breaking structure, or a closely-related
impostor, such as the radion of a warped extra-dimensional model.
Such a determination can only come by making improved measurements of
the particles properties and couplings to other particles.

One important observable that will help to establish the particle's
identity could be the production rate.  Unfortunately, the dominant
production mechanism for the Standard Model Higgs Boson, gluon fusion,
has a large theoretical uncertainty, of order $15\%$, even though is
has been computed to next-to-next-to-leading order in $\as$.  This
theoretical uncertainty receives two, roughly equal contributions: the
scale uncertainty in the partonic cross section and the uncertainty in
the values of the parton distributions.  The determination of parton
distributions will improve with further experimentation, but are
unlikely to be dramatically reduced.  The uncertainty in the partonic
cross-section, however, can be addressed by calculating ever-higher
orders in the expansion in $\as$.

The \nnlo\ calculation was completed in
2002~\cite{Harlander:2002wh,Anastasiou:2002yz,Ravindran:2003um} and is
now a mature result.  It is therefore time to address the extension to
next-to-next-to-next-to-leading order (\nnnlo).  Indeed, the process
has already started: The purely virtual corrections, the three-loop
corrections to $gg\to H$ were
computed~\cite{Lee:2010cga,Baikov:2009bg,Gehrmann:2010ue,Gehrmann:2010tu}
a couple of years ago; last year, the convolutions of \nnlo\ and
lower-order cross sections with the DGLAP splitting
functions~\cite{Hoschele:2012xc} were computed; and earlier this
year~\cite{Anastasiou:2013srw}, Anastasiou and collaborators reported
results for the first few terms in the threshold expansion of the
triple-real radiation contributions.  In this paper, I will present
the contributions from one-loop single-real-emission amplitudes.  Like
Ref.~\cite{Anastasiou:2013srw}, I too compute some of the terms which
appear by means of a threshold expansion.  However, by extending the
techniques established in
Refs.~\cite{Harlander:2002wh,Harlander:2002vv,Harlander:2003ai}, I am
able to map the expansion onto a set of basis functions consisting of
harmonic polylogarithms.  I am therefore able to report the complete
result as a Laurent series in the dimensional regularization parameter
($D=4-2\ep$) through order $\ep^{(1)}$.  The results for the
contributions at \nnnlo\ were recently computed, using very different
methods, in Ref.~\cite{Anastasiou:2013mca}.  After careful comparison,
we find that our results for the contribution to the inclusive cross
section agree completely.

The plan of this paper is as follows: In \Sec{sec:setup}, I will
describe the setup of the calculation: the structure of \nnnlo\
calculations; the effective Lagrangian and the resulting tree-level
and one-loop amplitudes; the calculation of loop master integrals and
renormalization.  In \Sec{sec:math} I will review the mathematical
structure of the functions I will be working with, namely harmonic
polylogarithms, (multiple) $\zeta$-functions and functions of uniform
transcendentality.   In \Sec{sec:methods}, I discuss the squaring of
the amplitudes and integration over phase space.  In particular, I
discuss the methods used to reduce and perform the phase space
integrals.  In \Sec{sec:results} I present results for the reduction
of phase space integrals to a set of master integrals and I present
the results for the partonic cross sections.  I present the \nlo\ and
\nnlo\ cross sections in closed form.  The expression for the \nnnlo\
partonic cross sections are very lengthy, so I present only the first
few terms in the threshold expansion.  The complete result (as a
Laurent expansion through order $\ep^{(1)}$) in terms of harmonic
polylogarithms is given in the supplementary material attached to this
article.  Finally, in \Sec{sec:conclude}, I present my
conclusions. 

\section{Setup of the Calculation}
\label{sec:setup}
\subsection{The Structure of \nnnlo\ Calculations}
\label{sec:nnnlo}
A perturbative calculation at \nnnlo\ contains many pieces.  It
contains virtual corrections ($gg\to H$) through three loops,
single-real-emission corrections through two loops,
double-real-emission corrections through one loop and
triple-real-emission at tree-level.  Each contribution lives in its
own phase space and must be computed separately from the others.  The
triple-real terms are computed from the squares of the tree-level
matrix elements.  The double-real terms from the squares of the
tree-level terms and the interference of the the tree-level amplitudes
with the one-loop amplitudes.  The single-real emission terms contain
the square of the tree-level amplitudes, the interference of
tree-level with one-loop amplitudes, the square of the one-loop
amplitudes and the interference of tree-level with the two-loop
amplitudes.  The purely virtual terms contain the squares of the
tree-level and one-loop amplitudes, the interference of the tree-level
amplitudes with one-, two- and three-loop amplitudes, and the
interference of one- and two-loop amplitudes.  In this paper, I will
focus on single-real emission corrections and restrict myself to terms
involving the one-loop amplitudes.  The contributions from the
interference of tree-level with two-loop amplitudes is left to future
consideration.

\subsection{The Effective Lagrangian}
\label{sec:ELag}
\begin{figure*}[ht]
\includegraphics[height=4.cm]{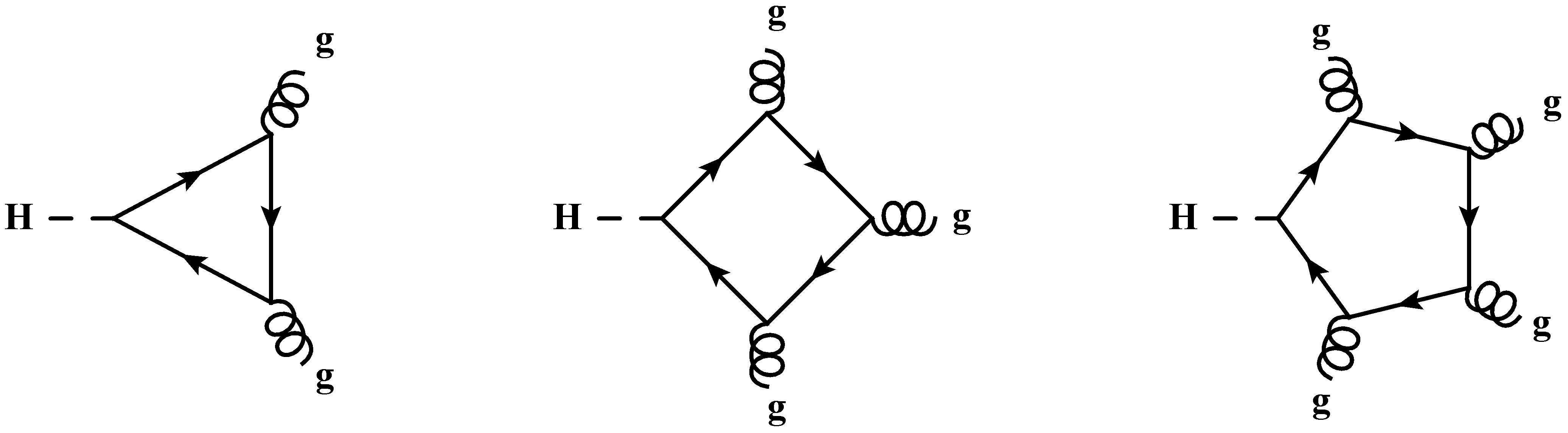}
\caption{\label{fig::toploops}Top loop diagrams
coupling gluons to the Higgs boson}
\end{figure*}

In the Standard Model, elementary particles obtain mass through their
couplings to the Higgs field.  Massless particles, like gluons and
photons, do not couple directly to the Higgs fields.  Instead, they
couple indirectly through massive particle loops.  In the limit that
all quarks except the top are massless, gluons couple to the Higgs
through top loops as shown in \fig{fig::toploops}, while photons
couple through both top and $W$ boson loops.  The light quarks
(treated as massless) couple to the Higgs fields through gluons and
photons feeding into massive particle loops.

\begin{figure*}[ht]
\includegraphics[height=4.cm]{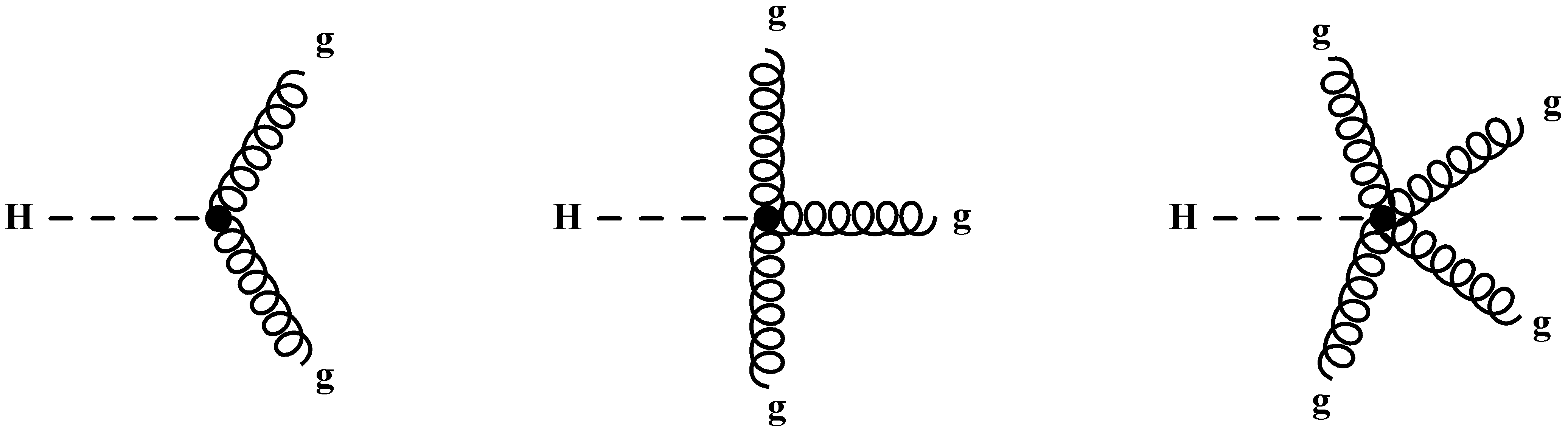}
\caption[]{\label{fig::effvert}Effective vertices coupling gluons to
the Higgs boson}
\end{figure*}

Since the $126$ GeV Higgs boson mass is far below the top threshold
($M_H \ll 2M_t$), one can integrate out the top quark and compute
amplitudes involving the Higgs field using \qcd\ with five active
flavors and the following effective
Lagrangian~\cite{Shifman:1979eb,Voloshin:1986tc,Vainshtein:1980ea} for
the Higgs-gluon interaction:
\begin{equation}
\begin{split}
{\cal L}_{\rm eff} 
&= -\frac{H}{4v} C_1^\bare(\alpha_s)\,{\cal O}_1^\bare
= -\frac{H}{4v} C_1(\alpha_s)\,{\cal O}_1\,,\qquad
{\cal O}_1 =  G^a_{\mu\nu} G^{a\,\mu\nu}\,,
\label{eqn::leff}
\end{split}
\end{equation}
where $v$ is the vacuum expectation value of the Higgs field $H$
($v\sim246\,$ GeV), $G^a_{\mu\nu}$ is the gluon field strength tensor
and the ${}^\bare$ superscripts represent bare quantities.  In the
approximation that all light flavors are massless, this effective
Lagrangian is renormalization group invariant, but the coefficient
function $C_1^\bare(\alpha_s)$ and the operator ${\cal O}_1^\bare$
must each be renormalized.  Using this effective Lagrangian, the top
quark loops of \fig{fig::toploops} are replaced by the effective
vertices shown in \fig{fig::effvert}.
The finite top mass corrections to the \nnlo\ result using this
effective Lagrangian are found to be very small for a Higgs mass near
$126$ GeV~\cite{Pak:2009dg,Harlander:2009my}.

The coefficient function $C_1(\alpha_s)$ contains the residual
logarithmic dependence on the top quark mass and has been computed up
to $\order{\alpha_s^4}$
\cite{Chetyrkin:1998un,Chetyrkin:1997iv,Schroder:2005hy,Chetyrkin:2005ia},
though for this calculation, one needs it only up to
$\order{\alpha_s^3}$~\cite{Chetyrkin:1998un,Chetyrkin:1997iv,Kramer:1998iq}.
In the modified minimal subtraction scheme ($\msbar$), the
renormalized coefficient function is:
\begin{widetext}
\begin{equation}
\begin{split}
  C_1(\alpha_s) &= -\frac{1}{3}\aspi\left\{ 1 + \frac{11}{4}\aspi +
  \aspi^2\left[\frac{2777}{288} + \frac{19}{16}\logmut{}
    + N_f\left(-\frac{67}{96} +
  \frac{1}{3}\logmut{}\right)\right]\right.\\
&\hskip 20pt
 + \aspi^3\left[-\frac{2761331}{41472} + \frac{897943}{9216}\zeta(3)
    + \frac{2417}{288}\logmut{}+\frac{209}{64}\logmut{2}\right.\\
&\hskip 50pt
    + N_f\left(\frac{58723}{20736} - \frac{110779}{13824}\zeta(3)
          + \frac{91}{54}\logmut{}
          + \frac{23}{32}\logmut{2}\right)\\
&\hskip 50pt\left.\left.
      + N_f^2\left(-\frac{6865}{31104} + \frac{77}{1728}\logmut{}
     - \frac{1}{18}\logmut{2}\right)\right]+
  \ldots\right\}\,,
\label{eqn::c1}
\end{split}
\end{equation}
\end{widetext}
where $\logmut{} = \ln(\mu^2/M_t^2)$, $\mu$ is the renormalization
scale and $M_t$ the on-shell top quark mass.  $\alpha_s \equiv
\alpha_s^{(5)}(\mu^2)$ is the $\msbar$ renormalized \qcd\ coupling
constant for five active flavors, and $N_f$ is five, the number of
massless flavors.

\subsection{$H\,g\,g\,g$ Amplitudes}
\label{sec:MEs}
\label{sec::HgggAmps}
The $H\,g\,g\,g$ amplitude can be written at any loop order in terms of
four linearly independent gauge invariant
tensors~\cite{Ellis:1988xu,Gehrmann:2011aa},
\begin{equation}
\begin{split}
 \label{eq::EHSvdBa}
   {\cal M}(H;1,2,3) &= \frac{g}{v}\,C_1(\alpha_s)\,f^{ijk}
   \ve^i_{1\mu}\ve^j_{2\nu}\ve^k_{3\rho}\sum_{n=0}^3A_n\,
  {\cal Y}_n^{\mu\nu\rho}.
\end{split}
\end{equation}
where $g$ is the \qcd\ coupling and $f^{ijk}$ are the structure
constants of $SU(N_c)$.  I adopt the following tensor definitions:
\begin{widetext}
\begin{equation}
\begin{split}
 \label{eq::Tensors}
   {\cal Y}_0^{\mu\nu\rho} =\ &
      \left(p_1^\nu g^{\rho\mu} - p_1^\rho g^{\mu\nu}\right)\,\frac{s_{23}}{2}
    + \left(p_2^\rho g^{\mu\nu} - p_2^\mu g^{\nu\rho}\right)\,\frac{s_{31}}{2}
    + \left(p_3^\mu g^{\nu\rho} - p_3^\nu g^{\rho\mu}\right)\,\frac{s_{12}}{2}
    + p_2^\mu p_3^\nu p_1^\rho - p_3^\mu p_1^\nu p_2^\rho\,,\\
   {\cal Y}_1^{\mu\nu\rho} =\ &
      p_2^\mu p_1^\nu p_1^\rho
    - p_2^\mu p_1^\nu p_2^\rho\,\frac{s_{31}}{s_{23}}
    - \frac{1}{2}p_1^\rho g^{\mu\nu}\,s_{12}
    + \frac{1}{2}p_2^\rho g^{\mu\nu}\frac{s_{31}\,s_{12}}{s_{23}}\,,\\
   {\cal Y}_2^{\mu\nu\rho} =\ &
      p_3^\mu p_3^\nu p_1^\rho
    - p_3^\mu p_1^\nu p_1^\rho\,\frac{s_{23}}{s_{12}}
    - \frac{1}{2}p_3^\nu g^{\mu\rho}\,s_{31}
    + \frac{1}{2}p_1^\nu g^{\mu\rho}\frac{s_{23}\,s_{31}}{s_{12}}\,,\\
   {\cal Y}_3^{\mu\nu\rho} =\ &
      p_2^\mu p_3^\nu p_2^\rho
    - p_3^\mu p_3^\nu p_2^\rho\,\frac{s_{12}}{s_{31}}
    - \frac{1}{2}p_2^\mu g^{\nu\rho}\,s_{23}
    + \frac{1}{2}p_3^\mu g^{\nu\rho}\frac{s_{12}\,s_{23}}{s_{31}}\,,\\
\end{split}
\end{equation}
\end{widetext}
The momenta are specified as if the process were $
H\,g_1\,g_2\,g_3\to\emptyset$.  Momentum conservation thus demands
that $p_H+p_1+p_2+p_3= 0$.

From these tensors, I can construct projectors to map the amplitudes
onto their tensor coefficients.
\begin{equation}
\begin{split}
{\cal P}_{{\cal Y}_{0}} =&
     \frac{D}{D-3}\frac{{\cal Y}_{0}}{s_{12}\,s_{23}\,s_{31}}
   - \frac{D-2}{D-3}\Lx
      \frac{{\cal Y}_{1}}{s_{31}\,s_{12}^{2}}
    + \frac{{\cal Y}_{2}}{s_{23}\,s_{31}^{2}}
    + \frac{{\cal Y}_{3}}{s_{12}\,s_{23}^{2}}
    \Rx\,,\\
{\cal P}_{{\cal Y}_{1}} =&
     \frac{D}{D-3}\frac{s_{23}\,{\cal Y}_{1}}{s_{31}\,s_{12}^{3}}
   - \frac{D-2}{D-3}\frac{{\cal Y}_{0}}{s_{31}\,s_{12}^{2}}
   + \frac{D-4}{D-3}\Lx
      \frac{{\cal Y}_{2}}{s_{12}\,s_{31}^{2}}
    + \frac{{\cal Y}_{3}}{s_{23}\,s_{12}^{2}}
    \Rx\,,\\
{\cal P}_{{\cal Y}_{2}} =&
     \frac{D}{D-3}\frac{s_{12}\,{\cal Y}_{2}}{s_{23}\,s_{31}^{3}}
   - \frac{D-2}{D-3}\frac{{\cal Y}_{0}}{s_{23}\,s_{31}^{2}}
   + \frac{D-4}{D-3}\Lx
      \frac{{\cal Y}_{3}}{s_{31}\,s_{23}^{2}}
    + \frac{{\cal Y}_{1}}{s_{12}\,s_{31}^{2}}
    \Rx\,,\\
{\cal P}_{{\cal Y}_{3}} =&
     \frac{D}{D-3}\frac{s_{31}\,{\cal Y}_{3}}{s_{12}\,s_{23}^{3}}
   - \frac{D-2}{D-3}\frac{{\cal Y}_{0}}{s_{12}\,s_{23}^{2}}
   + \frac{D-4}{D-3}\Lx
      \frac{{\cal Y}_{1}}{s_{23}\,s_{12}^{2}}
    + \frac{{\cal Y}_{2}}{s_{31}\,s_{23}^{2}}
    \Rx\,,
\end{split}
\end{equation}
where $D=4-2\ep$ is the dimensionality of space-time.

Each tensor coefficient has an expansion in $\alpha_s$ of the form:
\begin{equation}
  A_i = A_i^{(0)} + \aspi\,A_i^{(1)} + \aspi^2\,A_i^{(2)} + \ldots\,.
\end{equation}
I have computed the amplitudes in the following manner: the Feynman
diagrams were generated using \QGRAF~\cite{Nogueira:1993ex}; they were
contracted with the projectors onto the gauge-invariant tensors and
the Feynman rules were implemented using a
\FORM~\cite{Vermaseren:2000nd} program.  For the one-loop amplitudes,
the resulting expressions were reduced to loop master integrals with
the program \REDUZE~\cite{vonManteuffel:2012np}.  The reduced
expressions were put back into the FORM\ program and the master
integrals were evaluated to produce the final expressions.

At tree level, there are only four Feynman diagrams, and I find the
tree-level tensor coefficients to be
\begin{equation}
\begin{split}
\label{eq::Ctrees}
  A_0^{(0)} =& -\frac{2}{s_{12}} -\frac{2}{s_{23}} -\frac{2}{s_{31}},\qquad
            A_1^{(0)} = -\frac{2}{s_{31}},\qquad
            A_2^{(0)} = -\frac{2}{s_{23}},\qquad
            A_3^{(0)} = -\frac{2}{s_{12}}.
\end{split}
\end{equation}

\begin{figure}[ht]
a) \raise30pt\hbox{\includegraphics[width=4.cm]{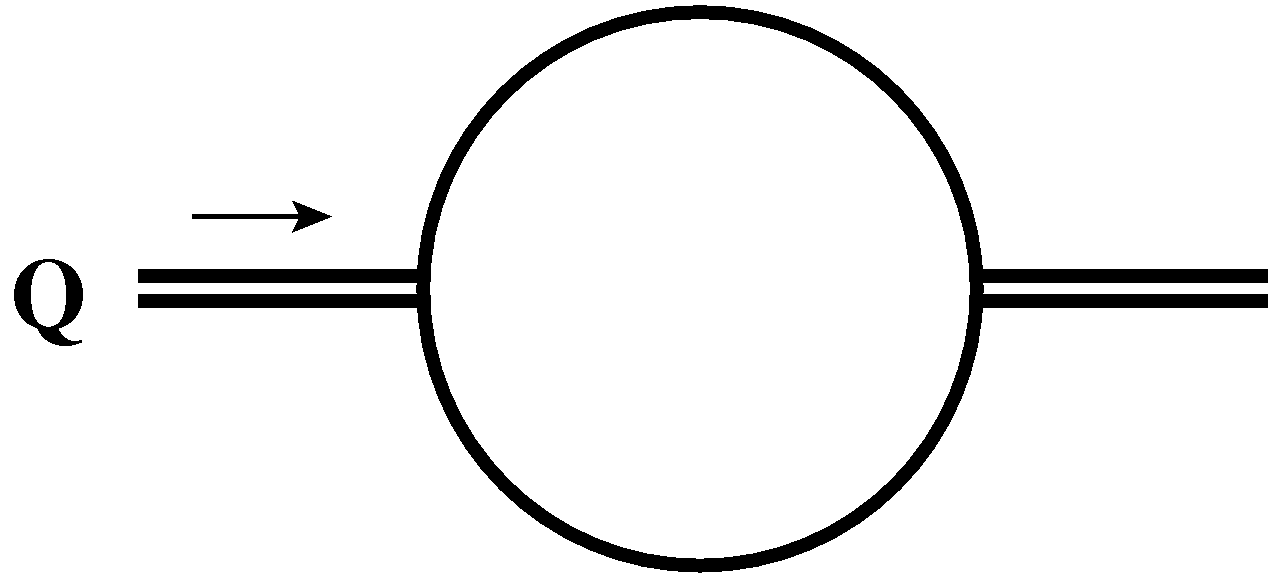}}\hskip 90pt
b) \includegraphics[width=4.cm]{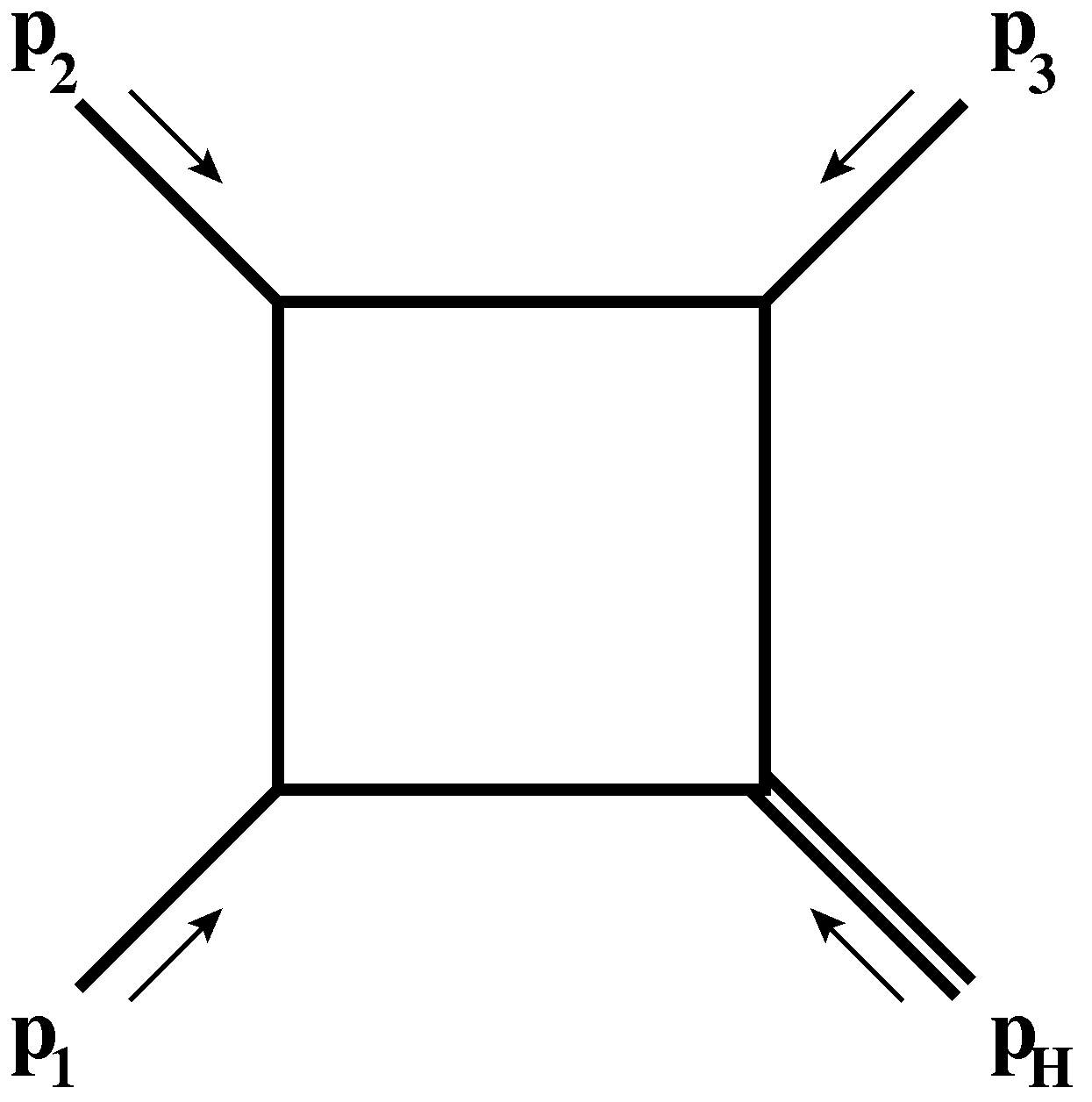}
\caption[]{\label{fig:masters}One-loop master integrals: a) The bubble
  diagram: ${\cal I}^{1}_{2}(Q^2)$, b) the box diagram with one massive leg:
${\cal I}^{1}_{4}(s_{12},s_{23};M_H^2)$.}
\end{figure}

There are only two master integrals involved in the one-loop
amplitude, the one-loop bubble, and the one-loop box with a single
massive external leg (see \fig{fig:masters}).

I find the one-loop tensor coefficients to be
\begin{equation}
\begin{split}
A_0^{(1)} = 4\,i\,\pi^2\,C_A&\left(a_{0,M}(s_{12},s_{23},s_{31})
        \,{\cal I}^{(1)}_2(M_H^2)
   + a_{0,s}(s_{12},s_{23},s_{31})\,{\cal I}^{(1)}_2(s_{12})
   + a_{0,s}(s_{23},s_{31},s_{12})\,{\cal I}^{(1)}_2(s_{23})\right.\\
  &\left.
   + a_{0,s}(s_{31},s_{12},s_{23})\,{\cal I}^{(1)}_2(s_{31})
   + \alpha_0(s_{12},s_{23},s_{31})\,{\cal I}^{(1)}_4(s_{12},s_{23};M_H^2)\right.\\
  &\left.
   + \alpha_0(s_{23},s_{31},s_{12})\,{\cal I}^{(1)}_4(s_{23},s_{31};M_H^2)
   + \alpha_0(s_{31},s_{12},s_{23})\,{\cal I}^{(1)}_4(s_{31},s_{12};M_H^2)\right)\,,\\
A_1^{(1)} = 4\,i\,\pi^2\,C_A&\left(a_{1,M}(s_{12},s_{23},s_{31})
        \,{\cal I}^{(1)}_2(M_H^2)
   + a_{1,s_{12}}(s_{12},s_{23},s_{31})\,{\cal I}^{(1)}_2(s_{12})
   + a_{1,s_{23}}(s_{12},s_{23},s_{31})\,{\cal I}^{(1)}_2(s_{23})\right.\\
  &\left.
   + a_{1,s_{31}}(s_{12},s_{23},s_{31})\,{\cal I}^{(1)}_2(s_{31})
   + \alpha_{1,2}(s_{12},s_{23},s_{31})\,{\cal I}^{(1)}_4(s_{12},s_{23};M_H^2)\right.\\
  &\left.
   + \alpha_{1,3}(s_{12},s_{23},s_{31})\,{\cal I}^{(1)}_4(s_{23},s_{31};M_H^2)
   + \alpha_{1,1}(s_{12},s_{23},s_{31})\,{\cal I}^{(1)}_4(s_{31},s_{12};M_H^2)\right)\,,
\end{split}
\end{equation}
where $C_A=N_c$ is the Casimir operator for the adjoint representation
and
\begin{equation}
\begin{split}
  a_{0,M}(s_{12},s_{23},s_{31}) &= \left[-(D-2)\,M_H^2\,\frac{1}{s_{12}^2}
     +(D-4)\,\frac{s_{12}}{s_{23}\,s_{31}}
     +\left(4\,\frac{D-4}{D-2}+12\,\frac{D-3}{D-4}\right)
     \frac{1}{s_{12}}\right.\\
   &\left.\hskip-50pt
      - 2\,\frac{(D-6)(D-4)}{D-2}\frac{1}{s_{23}+s_{31}}
     + 4\,\frac{D-4}{D-2}\frac{s_{12}}{(s_{23}+s_{31})^2}\right]
    +\left[\cdots\vp{\frac{1}{s_{12}}}
         \right]_{s_{12}\to s_{23}\to s_{31}\to s_{12}}
    +\left[\cdots\vp{\frac{1}{s_{12}}}
         \right]_{s_{12}\to s_{31}\to s_{23}\to s_{12}}\,,\\
  a_{0,s}(s_{12},s_{23},s_{31}) &= (D-2)\,M_H^2\,\left(
       \frac{1}{s_{23}^2}+\frac{1}{s_{31}^2}\right)
      -(D-4)\,\frac{s_{12}}{s_{23}\,s_{31}}
      - \frac{(D-2)^2}{D-4}\,\left(\frac{1}{s_{23}}
           +\frac{1}{s_{31}}\right)\\
  &    - 4\,\frac{(D-3)}{D-4}\frac{1}{s_{12}}
       - 4\,\frac{(D-4)}{D-2}\frac{M_H^2}{(s_{23}+s_{31})^2}\,,\\
  \alpha_{0}(s_{12},s_{23},s_{31}) &=
     \frac{1}{4}\frac{(D-2)(D-4)}{D-3}\frac{s_{12}\,s_{23}\,M_H^2}{s_{31}^2}
     - \frac{s_{12}\,s_{23} + s_{23}\,s_{31} + s_{31}\,s_{12}}{s_{31}}\,,
\end{split}
\end{equation}
and
\begin{equation}
\begin{split}
  a_{1,M}(s_{12},s_{23},s_{31}) &= D\,\frac{s_{23}\,M_H^2}{s_{12}^3}
     -\frac{(D-4)(3\,D-16)}{D-2}\frac{1}{s_{31}}
     + (D-4)\frac{(s_{23}+s_{31})^2\,M_H^2}{s_{12}\,s_{23}\,s_{31}^2}
     \\
  &  - \frac{D\,(D-4)}{D-2}\frac{M_H^4+s_{12}^2}{s_{12}^2\,s_{31}}
     - 2\,\frac{(D-4)^2}{D-2}\frac{1}{s_{12}+s_{23}}
     + 4\,\frac{D-4}{D-2}\frac{M_H^2}{(s_{12}+s_{23})^2}\\
  &  - 2\,\frac{(D-4)^2}{D-2}\frac{s_{23}}{s_{31}\,(s_{12}+s_{31})}
     + 4\,\frac{D-4}{D-2}\frac{s_{23}\,M_H^2}{s_{31}\,(s_{12}+s_{31})^2}
     + 12\,\frac{D-3}{D-4}\frac{1}{s_{31}}\,,\\
  a_{1,s_{12}}(s_{12},s_{23},s_{31}) &=-(D-4)\frac{M_H^2}{s_{12}\,s_{23}}
         \left(1 + \frac{s_{23}^2}{s_{31}^2}\right)
     + \frac{(D-4)^2}{D-2}\frac{M_H^2}{s_{12}\,s_{31}}\\
  &- \frac{(D-4)(D-2-2\,N_f/C_A)}{(D-1)(D-2)}\frac{s_{23}}{s_{12}^2}
   - 4\,\frac{D-3}{D-4}\frac{1}{s_{31}}\,,\\
  a_{1,s_{23}}(s_{12},s_{23},s_{31}) &=
     - (D-4)\frac{s_{23}\,M_H^2}{s_{12}\,s_{31}^2}
     - D\,\frac{s_{23}\,M_H^2}{s_{12}^3}
     + \frac{D\,(D-4)}{D-2}\frac{s_{23}\,M_H^2}{s_{12}^2\,s_{31}}\\
  &  - 4\,\frac{D-4}{D-2}\frac{s_{23}\,M_H^2}{s_{31}(s_{12}+s_{31})^2}
     - 4\,\frac{D-3}{D-4}\frac{1}{s_{31}}\,,\\
  a_{1,s_{31}}(s_{12},s_{23},s_{31}) &=
     - (D-4)\,\frac{M_H^2}{s_{12}\,s_{23}}
     - D\,\frac{s_{23}\,M_H^2}{s_{12}^3}
     + \frac{D\,(D-4)}{D-2}\frac{M_H^2}{s_{12}^2}\\
  &  
     - 4\,\frac{D-4}{D-2}\frac{M_H^2}{(s_{12}+s_{23})^2}
     - 4\frac{D-3}{D-4}\frac{1}{s_{31}}\,,\\
  \alpha_{1,1}(s_{12},s_{23},s_{31}) &= s_{12}
     + \frac{1}{4}\frac{(D-4)^2}{D-3}
         \frac{s_{31}\,M_H^2}{s_{23}}\,,\\  
  \alpha_{1,2}(s_{12},s_{23},s_{31}) &= 
     \frac{s_{12}\,s_{23}}{s_{31}}
     + \frac{1}{4}\frac{(D-4)^2}{D-3}
         \frac{s_{23}^2\,M_H^2}{s_{31}^2}\,,\\ 
  \alpha_{1,3}(s_{12},s_{23},s_{31}) &= 
     s_{23} + \frac{1}{4}\frac{D\,(D-4)}{D-3}
         \frac{s_{31}\,s_{23}^2\,M_H^2}{s_{12}^3}\,.
\end{split}
\end{equation}
The tensor coefficients $A_2^{(1)}$ and $A_3^{(1)}$ are given by
permutations of the invariants in $A_1^{(1)}$:
\begin{equation}
\begin{split}
  A_2^{(1)} &= \left.A_1^{(1)}
     \right|_{s_{12}\to s_{31}\to s_{23}\to s_{12}}\,,
  \hskip 40pt
  A_3^{(1)} = \left.A_1^{(1)}
     \right|_{s_{12}\to s_{23}\to s_{31}\to s_{12}}\,.
\end{split}
\end{equation}

\subsection{$H\,q\,\overline{q}\,g$ Amplitudes}
The $H\,q\,\overline{q}\,g$ amplitudes can be written in terms of only
two gauge invariant tensor structures,
\begin{equation}
\begin{split}
  \label{eq::Mgqq}
   {\cal M}(H;g,q, \overline{q}) =\ & 
      i\frac{g}{v}\,C_1(\alpha_s)\Lx T^g\Rx^{\bar\imath}_j\ve_\mu(p_g)
      \Lx B_1\,{\cal X}^\mu_1 + B_2\,{\cal X}^\mu_2\Rx\,,
\end{split}
\end{equation}
where $ T^g$ is a generator of the fundamental
representation of $SU(N_c)$ the tensors are given by~\cite{Gehrmann:2011aa}
\begin{equation}
\begin{split}
  {\cal X}_{1}^{\mu} &= p_{\overline{q}}^{\mu}
    \overline{u}(p_q)\,\slashed{p}_g\,v(p_{\overline{q}})
   - \frac{s_{\overline{q}g}}{2}
    \overline{u}(p_q)\,\gamma^\mu\,v(p_{\overline{q}})\,,\\
  {\cal X}_{2}^{\mu} &= p_{q}^{\mu}
    \overline{u}(p_q)\,\slashed{p}_g\,v(p_{\overline{q}})
   - \frac{s_{gq}}{2}
    \overline{u}(p_q)\,\gamma^\mu\,v(p_{\overline{q}})\,,
\end{split}
\end{equation}
and the projectors onto these tensors are
\begin{equation}
\begin{split}
{\cal P}_{{\cal X}_{1}} =& \frac{D-2}{D-3}
     \frac{{\cal X}_{1}^{\dagger}}{2\,s_{q\overline{q}}\,s_{\overline{q}g}^{2}}
   - \frac{D-4}{D-3}\frac{{\cal X}_{2}^{\dagger}}
       {2\,s_{q\overline{q}}\,s_{gq}\,s_{\overline{q}g}}\,,\\
{\cal P}_{{\cal X}_{2}} =& \frac{D-2}{D-3}
     \frac{{\cal X}_{2}^{\dagger}}{2\,s_{q\overline{q}}\,s_{gq}^{2}}
   - \frac{D-4}{D-3}\frac{{\cal X}_{1}^{\dagger}}
       {2\,s_{q\overline{q}}\,s_{gq}\,s_{\overline{q}g}}\,.\\
\end{split}
\end{equation}
These tensor coefficients also have expansions in $\alpha_s$:
\begin{equation}
  B_i = B_i^{(0)} + \aspi\,B_i^{(1)} + \aspi^2\,B_i^{(2)} + \ldots\,.
\end{equation}
The calculation proceeds through the same chain of \QGRAF, \FORM, and
\REDUZE\ programs as before.  The tree-level coefficients are:
\begin{equation}
  B_1^{(0)} = B_2^{(0)} = \frac{1}{s_{q\overline{q}}}\,,
\end{equation}
and the one-loop coefficients $B_i^{(1)}$ involve the same set of
master integrals as the $A_i^{(1)}$:
\begin{equation}
\begin{split}
B_1^{(1)} = -4\,i\,\pi^2\,C_A&\left(b_{1,M}(s_{q\overline{q}},
        s_{gq},s_{\overline{q}g})\,{\cal I}^{(1)}_2(M_H^2)
   + b_{1,s_{q\overline{q}}}(s_{q\overline{q}},s_{gq},
        s_{\overline{q}g})\,{\cal I}^{(1)}_2(s_{q\overline{q}})
   + b_{1,s_{gq}}(s_{q\overline{q}},s_{gq},
        s_{\overline{q}g})\,{\cal I}^{(1)}_2(s_{gq})\right.\\
  &\left.
   + b_{1,s_{\overline{q}g}}(s_{q\overline{q}},s_{gq},
        s_{\overline{q}g})\,{\cal I}^{(1)}_2(s_{\overline{q}g})
   + \beta_{1,\overline{q}}(s_{q\overline{q}},s_{gq},
        s_{\overline{q}g})\,{\cal I}^{(1)}_4(s_{q\overline{q}},s_{gq};M_H^2)\right.\\
  &\left.
   + \beta_{1,g}(s_{q\overline{q}},s_{gq},
        s_{\overline{q}g})\,{\cal I}^{(1)}_4(s_{gq},s_{\overline{q}g};M_H^2)
   + \beta_{1,q}(s_{q\overline{q}},s_{gq},
        s_{\overline{q}g})\,{\cal I}^{(1)}_4(s_{\overline{q}g},s_{q\overline{q}};M_H^2)\right)\,,\\
\end{split}
\end{equation}
where
\begin{equation}
\begin{split}
b_{1,M}(s_{q\overline{q}},s_{gq},s_{\overline{q}g}) &= 
    \frac{M_H^2}{s_{\overline{q}g}}\Lx\frac{D-4}{s_{q\overline{q}}} + \frac{D-4}{s_{gq}}
           - \frac{D-2}{s_{\overline{q}g}}\Rx + \frac{2}{D-4}\Lx1 +
    2(D-3)\frac{C_F}{C_A}\Rx\frac{1}{s_{q\overline{q}}}\\
 & - \frac{D^2-10\,D+20}{D-2}\frac{1}{s_{q\overline{q}}}
   - \frac{(D-4)\,M_H^2}{s_{q\overline{q}}\,(s_{q\overline{q}}+s_{\overline{q}g})}
   - \frac{(D-4)^2}{D-2}\frac{1}{s_{gq}+s_{\overline{q}g}}
   + 2\frac{D-4}{D-2}\frac{M_H^2}{(s_{gq}+s_{\overline{q}g})^2}\,,\\
b_{1,s_{q\overline{q}}}(s_{q\overline{q}},s_{gq},s_{\overline{q}g}) &= 
   \frac{D-2}{2}\frac{M_H^2}{s_{\overline{q}g}^2}
   - \frac{D-4}{2}\frac{M_H^2}{s_{gq}\,s_{\overline{q}g}}
   - \frac{1}{D-4}\Lx\frac{D^2-4\,D+12}{2} - \frac{C_F}{C_A}\Lx
        D^2-7\,D+16\Rx\Rx\frac{1}{s_{q\overline{q}}}\\
 & - \frac{1}{2}\frac{D-2}{D-1}\Lx1+2\frac{N_f}{C_A}\Rx\frac{1}{s_{q\overline{q}}}
   - 2\frac{D-4}{D-2}\frac{M_H^2}{(s_{gq}+s_{\overline{q}g})^2}\,,\\
b_{1,s_{gq}}(s_{q\overline{q}},s_{gq},s_{\overline{q}g}) &= 
   - \frac{D-4}{2}\frac{M_H^2}{s_{q\overline{q}}}\Lx
       \frac{1}{s_{\overline{q}g}} - 2\frac{C_F}{C_A}\frac{1}{s_{q\overline{q}}+s_{\overline{q}g}}\Rx
   + \frac{D-2}{2}\frac{M_H^2}{s_{\overline{q}g}^2}
   + 2\frac{D-3}{D-4}\Lx1-2\frac{C_F}{C_A}\Rx\frac{1}{s_{q\overline{q}}}\,,\\
b_{1,s_{\overline{q}g}}(s_{q\overline{q}},s_{gq},s_{\overline{q}g}) &= 
   - \frac{D-4}{2}\frac{1}{s_{\overline{q}g}}\Lx
           \Lx1-2\frac{C_F}{C_A}\Rx\frac{M_H^2}{s_{q\overline{q}}}
         + \frac{s_{q\overline{q}}+s_{\overline{q}g}}{s_{gq}}
         + \frac{C_F}{C_A}\Rx
   + 2\frac{D-3}{D-4}\Lx1-2\frac{C_F}{C_A}\Rx\frac{1}{s_{q\overline{q}}}\,,
\end{split}
\end{equation}
and
\begin{equation}
\begin{split}
\beta_{1,\overline{q}}(s_{q\overline{q}},s_{gq},s_{\overline{q}g}) &=
    - \frac{1}{2}\,s_{gq}
    + \frac{1}{8}\frac{(D-2)(D-4)}{D-3}
           \frac{s_{q\overline{q}}\,s_{gq}\,M_H^2}{s_{\overline{q}g}^2}\,,\\
\beta_{1,g}(s_{q\overline{q}},s_{gq},s_{\overline{q}g}) &=
      \Lx1-2\frac{C_F}{C_A}\Rx\frac{s_{gq}}{s_{q\overline{q}}}\Lx\frac{1}{2}\,s_{\overline{q}g}
        - \frac{1}{8}\frac{(D-4)^2}{D-3}\,M_H^2\Rx\,,\\
\beta_{1,q}(s_{q\overline{q}},s_{gq},s_{\overline{q}g}) &=
    - \frac{1}{2}\,s_{\overline{q}g}
    - \frac{1}{8}\frac{(D-4)^2}{D-3}\frac{s_{q\overline{q}}\,M_H^2}{s_{gq}}\,.
\end{split}
\end{equation}
$C_F = (N_c^2-1)/2/N_c$ is the Casimir operator of the fundamental
representation.  The other tensor coefficient, $B_2^{(1)}$ is given by
\begin{equation}
  B_2^{(1)}  = \left.B_1^{(1)}\right|_{s_{\overline{q}g}\leftrightarrow s_{gq}}
\end{equation}
These amplitudes describe all scattering configurations, $q\,\overline{q}\to
H\,g$, $g\,q\to H\,q$, etc., so long as the incoming and outgoing
momenta are correctly identified.

\subsection{Loop Master Integrals}
The loop master integrals that appear in these amplitudes are known in closed
form and the amplitudes are therefore known to all orders in the
dimensional regularization parameter $\ep$.  Working in the production
kinematics, where $s_{12} > 0$, $s_{23}, s_{31} < 0$
\begin{equation}
\begin{split}
\label{eqn:oneloopmasters}
{\cal I}^{(1)}_2(Q^2) &= 
    \frac{i\,\cg}{\ep\,(1-2\,\ep)}\left(\frac{\mu^2}{-Q^2}\right)^\ep\\
{\cal I}^{(1)}_4(s_{12},s_{23};M_H^2) &=
    \frac{2\,i\cg}{s_{12}\,s_{23}}\frac{1}{\ep^2}
    \left[\left(\frac{\mu^2}{-s_{12}}\right)^\ep
          \hypgeo{1}{-\ep}{1-\ep}{-\sos{31}{23}}\right.\\
   &\left.
        + \left(\frac{\mu^2}{-s_{23}}\right)^\ep
          \hypgeo{1}{-\ep}{1-\ep}{-\sos{31}{12}}
        - \left(\frac{\mu^2}{-M_H^2}\right)^\ep
          \hypgeo{1}{-\ep}{1-\ep}{-\frac{M_H^2\,s_{31}}{s_{12}\,s_{23}}}
    \right]\\
{\cal I}^{(1)}_4(s_{23},s_{31};M_H^2) &=
    \frac{2\,i\,\cg}{s_{23}\,s_{31}}\frac{1}{\ep^2}
    \left[\left(\frac{\mu^2}{-s_{12}}\right)^{-\ep}
      \left(\frac{\mu^2}{-s_{23}}\right)^\ep\left(\frac{\mu^2}{-s_{31}}\right)^\ep
      \Gamma(1-\ep)\Gamma(1+\ep)\right.\\
   &+ \left(\frac{\mu^2}{-s_{23}}\right)^\ep\Lx1-
          \hypgeo{1}{\ep}{1+\ep}{-\sos{31}{12}}\Rx
        + \left(\frac{\mu^2}{-s_{31}}\right)^\ep\Lx1-
          \hypgeo{1}{\ep}{1+\ep}{-\sos{23}{12}}\Rx\\
   &\left.
        - \left(\frac{\mu^2}{-M_H^2}\right)^\ep\Lx1-
          \hypgeo{1}{\ep}{1+\ep}{-\frac{s_{23}\,s_{31}}{s_{12}\,M_H^2}}\Rx
    \right]\,,\\
\end{split}
\end{equation}
where
\begin{equation}
\begin{split}
 \label{eq::cgamma}\cg = \frac{\Gamma(1+\ep)\Gamma^2(1-\ep)}
    {\Lx4\pi\Rx^{2-\ep}\Gamma(1-2\ep)}.
\end{split}
\end{equation}
The master integrals have been expressed in such a way that imaginary
parts come only from terms like
$\left(\frac{\mu^2}{-M_H^2}\right)^\ep$, when the kinematic invariant
is positive.  The correct analytic continuation of these terms is
given by the ``$i\epsilon$'' prescription of the Feynman propagator,
($-s_{ij} \to -(s_{ij}+i\epsilon$).  Note that the expression for the
box integral takes a different form when the incoming legs are
adjacent to one another (first form), so that one of the two-particle
invariants entering the diagram is time-like and when they are not
(second form) so that both two-particle invariants entering the
diagram are space-like.

The arguments of the hypergeometric functions have been arranged so
that the functions are real-valued and well-behaved throughout the
kinematic range.  Logarithmic singularities in the hypergeometrics,
resulting from collinear emission, occur only at boundary points and
are integrable.

\subsection{Renormalization}
The renormalization of ultraviolet divergences is performed in the
$\msbar$ scheme.  The bare \qcd\ coupling, $\asbare$ is replaced with
the renormalized coupling $\amsbar(\mu^2)$, evaluated at the
renormalization scale $\mu^2$.
\begin{equation}
   \asbare = \Lx\frac{\mu^2\,e^{\gamma_E}}{4\,\pi}\Rx^\ep
   Z_{\amsbar}\,\amsbar(\mu^2)
\end{equation}
The structure of the renormalization constant $Z_{\amsbar}$ is
determined entirely by its lowest order ($1/\ep$) poles, which in
turn define the \qcd\ $\beta$-function.
\begin{equation}
\begin{split}
\betabar{}(\amsbar) = \mu^2\frac{d}{d\,\mu^2}\frac{\amsbar}{\pi}
   &= -\ep\frac{\amsbar}{\pi}\Lx1 + \frac{\amsbar}{Z_{\amsbar}}
    \frac{\partial Z_{\amsbar}}{\partial\amsbar}\Rx^{-1}
   = -\ep\frac{\amsbar}{\pi} - \sum_{n=0}^\infty\,\betabar{n}\,\amsbarpi^{n+2}\,.
\label{eqn:cdrbetadef}
\end{split}
\end{equation}
With this normalization, the first two coefficients of the
$\beta$-function are:
\begin{equation}
 \betabar{0} = \frac{11}{12}C_A - \frac{1}{6}N_f\,,\qquad\qquad
    \betabar{1} = \frac{17}{24}C_A^2 - \frac{5}{24}C_A\,N_f
          - \frac{1}{8}C_F\,N_f\,,
\end{equation}

The composite operator of the effective Lagrangian (\eqn{eqn::leff})
renormalizes as
\begin{equation}
{\cal O}_1^\bare = Z_{{\cal O}_1}{\cal O}_1\,,
\end{equation}
where~\cite{Chetyrkin:1996ke}
\begin{equation}
Z_{{\cal O}_1} = \Lx1+\frac{\amsbar}{Z_{\amsbar}}
    \frac{\partial Z_{\amsbar}}{\partial\amsbar}\Rx
   = \LB1 + \sum_{n=0}^\infty\,\betabar{n}\,\amsbarpi^{n+1}\RB^{-1}
\end{equation}
The Wilson coefficient, $C_1$, renormalizes in the exact opposite
fashion as the operator ${\cal O}_1$,
\begin{equation}
C_1^\bare = Z^{-1}_{{\cal O}_1}C_1\,.
\end{equation}
The value for $C_1$ given in \eqn{eqn::c1} is for the renormalized
Wilson coefficient.

\section{Mathematical Framework}
\label{sec:math}
Performing this calculation relies on taking advantage of the special
properties of the mathematical functions that appear in Feynman
integrals.  In particular, I make use of the harmonic polylogarithms
and the (multiple) $\zeta$-function.  These functions are closely
related, as I shall briefly describe below.

\subsection{Harmonic Polylogarithms}
The results of the calculations presented in this paper are
conveniently expressed in terms of harmonic polylogarithms.  The
mathematical properties of harmonic polylogarithms (HPL) have been
discussed extensively in the literature\
\cite{Remiddi:1999ew,Gehrmann:2000zt,Gehrmann:2001ck,Maitre:2005uu},
but I briefly review their definition and some important properties.

The standard harmonic polylogarithms are defined in terms of three
weight functions, $f_{+1}$, $f_0$, and $f_{-1}$:
\begin{equation}
  f_{+1}(x) = \frac{1}{1-x}\,,\qquad f_0(x) = \frac{1}{x}\,,\qquad
   f_{-1}(x) = \frac{1}{1+x}
\end{equation}
The weight-one HPLs are defined by
\begin{equation}
  H(0;x) = \ln\,x\,,\qquad H(\pm1;x) = \int_0^x\,dz\,f_{\pm1}(z)\,.
\end{equation}
Higher-weight HPLs are defined by iterated integrations against the
weight functions:
\begin{equation}
\label{eqn:HPLint}
  H(w_n,w_{n-1},\ldots,w_0;x)
     = \int_0^x\,dz\,f_{w_n}(z)\,H(w_{n-1},\ldots,w_0;z)\,,
\end{equation}
Clearly, the derivatives of HPLs involve the same weight functions,
\begin{equation}
  \frac{d}{dz}H(w_n,w_{n-1},\ldots,w_0;x) =
         f_{w_n}(z)\,H(w_{n-1},\ldots,w_0;z)\,.
\end{equation}
The HPLs include the classic polylogarithms, $\Li{n}(x)$ as special
cases.  For example, $\Li{1}(x) = H(1;x)$, $\Li{2}(x) = H(0,1;x)$,
 $\Li{3}(x) = H(0,0,1;x)$, etc.

There is a commonly-used shorthand notation for the weight vector
$\vec{w}$: whenever a weight $0$ is to the left of a non-zero weight,
the zero is omitted and the non-zero weight is increased in magnitude
by $1$.  So, $H(0,1;x)\to H(2:x)$, and $H(0,-1,0,0,1;x)\to
H(-2,3:x)$.

The HPLs are very versatile.  For example, it is relatively simple to
transform the argument of the HPLs to, for example, relate a function
$H(\vec{w};x)$ to a combination of functions $H(\vec{v},1-x)$.

Another important property is that the HPLs form a shuffle algebra, so
that
\begin{equation}
\label{eqn:shuffle}
  H(\vec{w}_1;x)\,H(\vec{w}_2;x) = \sum_{\vec{w}^\prime \in
    \vec{w}_1\shuf\vec{w}_2} H(\vec{w}\prime;x)\,,
\end{equation}
where $\vec{w}_1\shuf\vec{w}_2$ is the set of shuffles, or mergers of
the sequences $\vec{w}_1$ and $\vec{w}_2$ that preserve their relative
orderings.

The harmonic polylogarithms can be extended in various ways.  One is
to use different weight functions.  These additional weights can even
be related to kinematic
variables~\cite{Gehrmann:2000zt,Gehrmann:2001ck}.  In this work, it
will be convenient to introduce the weight function
\begin{equation}
  f_{+2}(x) = \frac{1}{2-x}\,,
\end{equation}
and the associated polylogarithms.  Using harmonic polylogarithms
derived from this weight function makes it confusing to try to use the
short-hand notation described above.  I therefore use it only when
working with standard HPLs and multiple $\zeta$-functions, and avoid
it when working with extended HPLs.

\subsection{Multiple $\zeta$-Functions}
The Multiple $\zeta$-function is a generalization of the Riemann
$\zeta$-function, defined by
\begin{equation}
\zeta(w_1,\ldots\,w_k) \equiv \sum_{n_1>n_2>\ldots>n_k}^\infty
    \frac{1}{n_1^{w_1}\cdots n_k^{w_k}}\,.
\end{equation}
When all weights $w_m$ are positive, these are sometimes called
multiple $\zeta$ values, or MZVs.  The multiple $\zeta$-functions
are, in some sense, the endpoints of the harmonic polylogarithms,
since
\begin{equation}
  H(\vec{w};1) = \zeta(\vec{w})\,,
\end{equation}
where $\vec{w}$ is written in the shorthand notation defined above.
If one deconstructs the weight vector into $0$'s, $+1$'s and $-1$'s, it
is clear that the multiple $\zeta$-functions share the shuffle algebra
of the harmonic polylogarithms.  This property allows one to derive
many relations involving the products and sums of the MZVs.

One important result is that at any rank $n$, the MZVs with weight
vectors containing only $2$'s and $3$'s form a basis for MZVs of that
rank~\cite{Hoffman:1467164,Blumlein:2009cf,Brown:2012XX,Zagier:2012XX}.
A consequence of this is that through rank $7$, one can replace this
basis with products of single (Riemann) $\zeta$ functions.  Not until
rank $8$ are there more elements in the basis ($\zeta(2,2,2,2)$,
$\zeta(3,3,2)$, $\zeta(3,2,3)$, $\zeta(2,3,3)$) than there are
independent single $\zeta$ products ($\zeta(8)$, $\zeta(5)\,\zeta(3)$,
$\zeta^2(3)\,\zeta(2)$).

\subsection{Functions of Uniform Transcendentality}
It is useful to define the concept of the degree of
transcendentality~\cite{Henn:2013pwa} ${\cal T}(f)$ of a function $f$
which, like the HPLs, is defined by iterated integration.  The degree
of transcendentality is simply the number of iterated integrals needed
to define the function.  Thus, the transcendentality of an HPL is
equal to the rank of its weight vector.  Transcendentality is also
assigned to numerical constants that are obtained at special values of
transcendental functions.  Thus $\zeta(5) = \Li{5}(1) =
H(5;1)$ is assigned ${\cal T}(\zeta(5)) = 5$.  The
transcendentality of products of functions is equal to the sum of the
transcendentalities of the two functions, ${\cal T}(f_1\,f_2) = {\cal
  T}(f_1) + {\cal T}(f_2)$.  This is consistent with the shuffle
operation where the product of functions of rank $r_1$ and $r_2$ is
expressed as a sum of functions of rank $r_1+r_2$.

A function is said to be a function of uniform
transcendentality~\cite{Henn:2013pwa} (FUT) if it is a sum of terms
which all have the same transcendentality.  A further refinement is to
define a {\it pure} function of uniform transcendentality (pFUT) as
one for which the degree of transcendentality is lowered by taking a
derivative, ${\cal T}(d\,f) = {\cal T}(f) -1$.  For instance, $f(x) =
x\,H(1;x)$ is not a pFUT because $df/dx = H(1;x) + x/(1-x)$ does not
have uniform transcendentality and thus is not an FUT, while $g(x) =
H(1,1;x) + H(0,1;x)$ is a pure function of uniform transcendentality
and $dg/dx = H(1;x)\,\Lx f_{1}(x) + f_{0}(x)\Rx$ is an FUT.

Typically, the functions that are encountered in performing
dimensionally regularized Feynman integrals are expressed as Laurent
expansions in the parameter $\ep$, where $D=4-2\ep$.  The concept of
transcendentality can by usefully applied to these functions by
assigning ${\cal T}(\ep)=-1$.  Simple examples of pure functions of
uniform transcendentality are
\begin{equation}
\begin{split}
\label{eqn:FUT1}
\Gamma(1-\ep) &= \exp\Lx\ep\,\gamma_E + \sum_{n=2}^{\infty}\frac{\ep^{n}\,\zeta(n)}{n}\Rx\\
\Lx\frac{\mu^2}{M_H^2}\Rx^\ep &= \sum_{n=0}^\infty
    \frac{\ep^n}{n!}\ln^{n}\frac{\mu^2}{M_H^2}
\end{split}
\end{equation}
where the Euler-Mascheroni constant, $\gamma_E \approx 0.577216$ is assigned
${\cal T}(\gamma_E) = 1$.  A more complicated example is the
hypergeometric function that appears in the one-loop box master
integrals (\eqn{eqn:oneloopmasters},
\begin{equation}
  \hypgeo{1}{-\ep}{1-\ep}{z} = 1 - \sum_{n=1}^\infty \ep^n\,\Li{n}(z)\,.
\end{equation}
Note that the one-loop bubble master integral, however, is not an FUT
because of the factor of $1/(1-2\,\ep)$.

\section{Methods}
\label{sec:methods}
\subsection{Squared amplitudes and Phase Space Integration}
\label{sec:SRE}
The partonic cross section is computed by squaring the production
amplitudes, averaging (summming) over initial (final) state colors and
spins, and integrating over phase space.
\begin{equation}
\sigma = \frac{1}{2\,s_{12}}\,d(LIPS)\,\frac{1}{\mathbb{S}}
  \sum_{\rm spin/color} \,\left|{\cal M}\right|^2\,,
\end{equation}
where the factor of $1/(2\,s_{12})$ is the flux factor, $d(LIPS)$
represents Lorentz invariant phase space and the factor $\mathbb{S}$
represents the averaging over initial state spins and colors.  The
matrix elements presented in the previous sections were written for
the kinematics $p_1+p_2+p_3+p_H\to\emptyset$.  To compute the cross
section, $p_3$ and $p_H$ must be crossed into the final state.  When
$p_3$ represents the momentum of a fermion, the squared matrix element
picks up an extra factor of $(-1)$ from Fermi-Dirac statistics.  For
the production process $p_1+p_2\to p_3+p_H$, the element of Lorentz
invariant phase space is
\begin{equation}
d(LIPS) = \frac{1}{8\,\pi}\Lx\frac{4\,\pi\,\mu^2}{s_{12}}\Rx^\ep\,\
   \frac{\Lx s_{23}\,s_{31}\Rx^\ep}{\Gamma(1-\ep)}\,ds_{23}\,.
\end{equation}
Defining $s_{12} = \hat{s}$ to be the parton CM energy squared, I
introduce the dimensionless parameters $x = M_H^2/\hat{s}$,
$\xb = 1-x$, and $y = \frac{1}{2}\Lx1-\cos\,\theta^*\Rx$, $\yb = 1-y$, where
$\theta^*$ is the scattering angle in the CM frame,
\begin{equation}
\begin{split}
  s_{12} &= \hat{s}\,,\hskip 50pt  M_H^2 = x\,\hat{s}\,,\\
  s_{23} &= \xb\,y\,\hat{s}\,,\hskip 37pt  s_{31} = \xb\,\yb\,\hat{s}\,.
\end{split}
\end{equation}
In terms of these variables, the element of phase space is
\begin{equation}
  d(LIPS) = \frac{1}{8\,\pi}\Lx\frac{4\,\pi^2\,\mu^2}{\hat{s}}\Rx^\ep
    \frac{1}{1-\ep}\,\xb^{1-2\ep}\,y^{-\ep}\,\yb^{-\ep}\,dy\,.
\end{equation}
$\xb$ is called the threshold parameter, and is a measure of excess or
kinetic energy in the scattering process, beyond that which is needed to
produce a Higgs boson at rest.  The kinematically available region in
$x$ and $y$ space is $M_H^2/s < x < 1$ and $0 < y < 1$, where
$s$ is the hadronic (not partonic) CM energy.  Clearly, $0 < \xb\ <
1-M_H^2/s$ and $0<\yb\ <1$.

In the virtual production process, $g\,g\to H$, there is no excess
energy and $\xb$ is constrained to be zero.  This constraint is
enforced by a $\delta$-function, $\delta(\xb)$, which arises from the
phase space element of the virtual process.  In a real emission
process, like that considered here, $\xb$ is allowed to vary
continuously between $0$ and $M_H^2/s$ and the terms in the cross
section are multiplied by powers (both integer and proportional to
$\ep$) of $\xb$.  The leading terms in $\xb$, associated with soft
emission, vary like $\xb^{-1+n\,\ep}$, and are singular at the
endpoint $\xb\to0$.  These soft terms are evaluated by expanding in
distributions of $\xb$,
\begin{equation}
\label{eqn:xbdistrib}
\xb^{-1+n\,\ep} = \frac{1}{n\,\ep}\delta(\xb)
     + {\cal D}^{n\,\ep}(\xb)
     = \frac{1}{n\,\ep}\delta(\xb)
     + \sum_{m=0}^\infty\frac{(n\,\ep)^m}{m!}\,{\cal D}_m(\xb)\,,
\end{equation}
where ${\cal D}_m(\xb)$ is a ``plus'' distribution defined as
\begin{equation}
\begin{split}
  {\cal D}_m(\xb) &= \LB\frac{\ln^m(\xb)}{\xb}\RB_+\,,\\
  \int_0^1 dx\,h(x)\,{\cal D}_m(\xb) &=
   \int_0^1 dx\,\Lx h(x)-h(1)\Rx\frac{\ln^m(\xb)}{\xb}\,,
\end{split}
\end{equation}
and ${\cal D}^{n\,\ep}(\xb)$ represents the whole tower of plus
distributions.  In this way, one obtains $\delta$-function terms to
add to those from the virtual corrections.

\subsection{Integration by Parts}
The partonic cross sections are given by integrals of the squared
matrix elements over the phase space.  This involves a great many
integrals of functions of varying complexity.  It is certainly
possible to simply attack the list of integrals, one-by-one, and solve
them by whatever means possible.  The magnitude of the problem can be
essentially cut in half by taking advantage of the symmetry in
exchanging $y\leftrightarrow\yb$, but this still leaves a large number
of integrals to be performed.

An elegant solution is suggested by the success of the
integration-by-parts method that has been applied to Feynman
integrals, allowing one to express a large set of integrals in terms
of a few ``master'' integrals.  Since loop integrals and phase space
integrals are intimately related through the Cutkosky relations, it is
no surprise that the same procedure can be applied to phase space
integrals.  An example of a phase space integral encountered in the
interference of tree-{} and one-loop amplitudes is
\begin{equation}
  I_{ex}(\xb) = \int_0^1 dy\,y^{-\ep}\,\yb^{-2\ep}\hypgeo{1}{-\ep}{1-\ep}{\xb\,y}\,.
\end{equation}
If I differentiate both sides of this equation by $y$, I obtain zero
on the left-hand side, since $I_{ex}(\xb)$ is not a function of $y$,
but when I carry the differential under the integral on the right-hand
side, I obtain a sum of different integrals.  Since the sum is equal
to zero, I derive non-trivial relations among various phase space
integrals.  In the example given above, I obtain
\begin{equation}
\begin{split}
0 &= \Lx1-2\,\ep\Rx\int_0^1 dy\,
     y^{-\ep}\,\yb^{-2\ep}\hypgeo{1}{-\ep}{1-\ep}{\xb\,y}\\
  &+2\,\ep\int_0^1 dy\,
     y^{-\ep}\,\yb^{-1-2\ep}\hypgeo{1}{-\ep}{1-\ep}{\xb\,y}\\
  &-\ep\int_0^1 dy\,
     y^{-\ep}\,\yb^{-2\ep}\Lx1-\xb\,y\Rx^{-1}
\end{split}
\end{equation}
As it turns out, two of the integrals on the right-hand side,
\begin{equation}
  \int_0^1 dy\, y^{-\ep}\,\yb^{-2\ep}\Lx1-\xb\,y\Rx^{-1} =
    \frac{\Gamma(1-\ep)\Gamma(1-2\,\ep)}{\xb\,\Gamma(1-3\,\ep)}
    \Lx-1 + \hypgeo{1}{-\ep}{1-3\,\ep}{\xb}\Rx\,,
\end{equation}
and
\begin{equation}
  \int_0^1 dy\, y^{-\ep}\,\yb^{-1-2\ep}\hypgeo{1}{-\ep}{1-\ep}{\xb\,y} =
    \frac{\Gamma(1-\ep)\Gamma(1-2\,\ep)}{(-2\,\ep)\Gamma(1-3\,\ep)}
    \hypgeo{1}{-\ep}{1-3\,\ep}{\xb}\,,
\end{equation}
are functions of uniform transcendentality. This makes them good
candidates to be chosen as master integrals, though as it turns out, I
have chosen other FUTs as masters.

\subsection{Threshold Expansion}
\label{sec:thresh}
Once the full set of integrals has been reduced to a few masters, one
must actually perform those integrals.  Some of the masters can be
integrated in closed form, but most of those that arise
from the squared one-loop amplitudes, cannot.  The technique by which
I will solve these integrals involves expansion of the integrands in
terms of the threshold parameter
$\xb$~\cite{Harlander:2002wh,Harlander:2002vv,Harlander:2003ai}.

The advantage of this approach is that the coefficient of each power
of $\xb$ consists of simple, often trivial, integrals over powers and
functions of $y$ and $\yb$ only.  The disadvantage is that the result
is a truncated series in $\xb$, not a set of functions in closed form.
This disadvantage, however, is essentially one of {\ae}sthetics.
Because the gluon luminosity spectrum is a fairly steeply falling
function, the Higgs production cross section is dominated by the
threshold region and so the first several terms in the $\xb$ expansion
give a good approximation to the physics.  This feature was
demonstrated explicitly in the first \nnlo\ calculation of Higgs
boson production~\cite{Harlander:2002wh}.

Nevertheless, even this disadvantage can be overcome if one has a
suitable ansatz for the basis of functions in which the closed-form
integrals would take values and if one can carry out the threshold
expansion to sufficiently high order that one can map the series
expansion onto the basis functions~\cite{Harlander:2002vv,Harlander:2003ai}.
At \nnlo, the author used the ansatz that the basis of functions
consisted of those functions which appeared in the ground-breaking
calculation of Drell-Yan production at \nnlo~\cite{Hamberg:1990np}.

In the present calculation, one does not have such guidance for how to
choose functions beyond rank three.  A logical choice would seem to be
the standard harmonic polylogarithms in $\xb$.  This, however, would
be incorrect!  Among the functions found in the \nnlo\ Drell-Yan
result are
\begin{equation}
  \Li{2}(-x) = - H(0,-1;x)\,, \qquad
  \Li{3}(-x) = - H(0,0,-1;x)\,.
\end{equation}
These functions can be expanded in $\xb$.  For example,
\begin{equation}
\label{eqn:inhomo}
  \Li{2}(-x) = - \frac{1}{2}\Li2(2\,\xb-\xb^2) + \Li2(\xb)
  - \frac{\zeta(2)}{2} + \ln(2)\Li1(\xb)
  - \Li1(\xb)\Li1\Lx\frac{\xb}{2}\Rx\,.
\end{equation}
All of the functions on the right hand side of this expression can be
readily expanded in $\xb\!$, but cannot be expressed as standard HPLs
of $\xb$.  A better ansatz is that the basis of functions consists of
standard HPLs of $x$, not $\xb$.  The problem with this ansatz,
however, is that the threshold expansion is in $\xb$, not $x$, and the
expansion in $\xb$ of HPLs in $x$ involves the appearance of
transcendental numbers like $\zeta(n)$ or $\ln(2)$ as in
\eqn{eqn:inhomo} above.  It turns out that the best basis of functions
consists of the generalized harmonic polylogarithms in $\xb$, where
the elements of the weight vector takes values from the set
$\{0,1,2\}$, rather than the standard $\{-1,0,1\}$.  These generalized
HPLs all expand homogeneously in $\xb$, without the appearance of
transcendental numbers.  Once the threshold expansion has been mapped
onto these functions, they can, in turn, be mapped back onto the
standard HPLs in $x$.  Thus, the final results of this calculation
will be expressed in terms of standard HPLs in $x$.

\subsection{Series Inversion}
The mapping of the threshold expansions onto basis functions is done
as follows.  For a set of $n$ basis functions, $G(w_i;\xb)$, (Note
that I use $G(\vec{w};x)$ to denote that I am using generalized rather
than standard HPLs) each function is expanded in powers of $\xb$ from
$\xb^0$ to $\xb^{n-1}$.  This statement assumes that the right-most
element of the weight vector is not equal to $0$.  Such terms would
contain factors of $\ln(\xb\!)$, which does not expand in powers of
$\xb$.  (There is no problem with eliminating these terms from the
basis since factors of $\ln(\xb\!)$ arise exclusively from terms like
$\xb^{n\,\ep}$, which appear explicitly in the phase space element and
have been factored out in the form of the loop master integrals given
in \eqn{eqn:oneloopmasters}.)  With this assumption, the HPLs can be
expanded as~\cite{Maitre:2005uu}
\begin{equation}
  G(\vec{w};\xb) = \sum_{i=0}^\infty\,\xb^{i}\ Z_i(\vec{w})\,.
\end{equation}
The coefficients $Z_i(\vec{w})$ can be determined using the definition
of the HPLs.
\begin{equation}
\label{eqn:zexp}
G(w_n,w_{n-1}\ldots,w_1;z) = \int_0^z dt\ f_{w_n}(t)\
   G(w_{n-1}\ldots,w_1;t) = \sum_{i=0}^\infty Z_i(w_{n-1}\ldots,w_1)
    \int_0^z\ f_{w_n}(t)\ t^i
\end{equation}
For $w_n$ taking values from the set $\{0,1,2\}$.
\begin{equation}
\begin{split}
\label{eqn:wts}
\int_0^z\ dt\ f_{0}(t)\ t^i &= \int_0^z\ \frac{dt}{t} t^i = \frac{z^i}{i}\,,\\
\int_0^z\ dt\ f_{1}(t)\ t^i &= \int_0^z\ \frac{dt}{1-t} t^i = 
    \sum_{j=i+1}^\infty \frac{z^j}{j}\,,\\
\int_0^z\ dt\ f_{2}(t)\ t^i &= \int_0^z\ \frac{dt}{2-t} t^i = 
    \sum_{j=i+1}^\infty \frac{z^j}{2^{j-i}j}\,.
\end{split}
\end{equation}
Combining \eqns{eqn:zexp}{eqn:wts}, I obtain starting values
\begin{equation} Z_j(1) = \frac{1}{j}\,, \qquad
   Z_j(2) = \frac{1}{(2)^j\,j}
\end{equation}
and the recursion relations:
\begin{equation}
\begin{split}
Z_j(0,\vec{w}) & = \frac{1}{j}\ Z_{j}(\vec{w})\,,\\
Z_j(1,\vec{w}) &= \frac{1}{j}\ \sum_{i=1}^{j-1}\ Z_{i}(\vec{w})\,,\\
Z_j(2,\vec{w}) &= \frac{1}{j}\ \sum_{i=1}^{j-1}\ \frac{1}{2^{j-i}}
        \ Z_{i}(\vec{w})\,.
\end{split}
\end{equation}

Once the basis functions have been expanded, one forms a matrix ${\mathbb
  M}$ of coefficients, where each column corresponds to a different
function, and each row to a different order in $\xb$.  This matrix is
inverted, to form ${\mathbb M}^{-1}$.  The solution to the integral
$I(\xb\!)$ is then found to be
\begin{equation}
  I(\xb\!) = \vec{f}\cdot {\mathbb M}^{-1}\cdot \vec{\imath}\,,
\end{equation}
where $\vec{f}$ is a row-vector of the basis functions, and
$\vec{\imath}$ is a column-vector consisting of the threshold
expansion coefficients of the integral $I(\xb\!)$.

Threshold expansion followed by series inversion is a very powerful
and versatile tool.  It can be used as a blunt instrument to invert
the threshold expansion of the entire partonic cross section.  This is
how it was used in the calculations of \nnlo\ Higgs cross
sections~\cite{Harlander:2002vv,Harlander:2003ai}.  When applied to
such complicated integrands, one needs not just the basis functions
discussed above, but also those basis functions weighted by various
powers of $\xb$.  Thus, while the inversion was performed using only
functions of rank $3$ or less (of which there are $40$ in total,
counting $1$ as a rank-$0$ function, and only $13$ which appear), we
needed a basis of $78$ functions.

The full power of the technique emerges, however, when it is applied
to a more controlled set of integrals.  As discussed above, I only
need to evaluate a relatively small number of master integrals.  The
rest are determined from the masters by algebraic relations.  If I
choose my master integrals to be pure functions of uniform
transcendentality, I significantly reduce the size of the basis needed
for inversion.  This is an important consideration because the number
of operations required for matrix inversion grows like $n^3$, where
$n$ is the size of the basis.  This $n^3$ growth in the number of
operations does not take into account the fact that the size of the
terms being manipulated also grows rapidly with $n$.  Thus, a
reduction in the size of the basis by a factor of $2$ makes the
problem of matrix inversion at least $10$ times simpler.  I find that
the most complicated integrals in this calculation require a basis of
only $48$ functions to extract the rank $5$ components.  In contrast,
to proceed by brute-force and compute the coefficients through rank
$5$ of the non-FUT integrals would require a basis of up to $325$
functions.

\section{Results}
\label{sec:results}
The first task is to compute the master integrals.

\subsection{Master Integrals at \nlo}
There is only one master integral that contributes to the integral of
the square of tree-level amplitudes over phase space.
\begin{equation}
\label{eqn:MINLO}
M_0 = \alpha\ep\int_0^1\ dy\,y^{-1+\alpha\,\ep}\ \yb^{\beta\,\ep} = 
\frac{\Gamma(1+\alpha\,\ep)\Gamma(1+\beta\,\ep)}
      {\Gamma(1+(\alpha+\beta)\ep)}
\end{equation}
For this integral, integration by parts does not yield any identities
that are not equivalent to $\Gamma(\alpha+1) = \alpha\,\Gamma(\alpha)$.

\subsection{Master Integrals at \nnlo}
Applying the integration-by-parts technique to the integrals that
appear in the interference of tree-{} and one-loop amplitudes, I find
that there are only five new master integrals.  All five can be
evaluated in closed form, meaning that the entire contribution of
single-real-emission at \nnlo\ can be evaluated to all orders in
$\ep$.  The master integrals are:
\begin{equation}
\begin{split}
\label{eqn:MINNLO}
M_1(\alpha,\beta) = \alpha\,\ep\,&\int_0^1\ dy\,y^{-1+\alpha\,\ep}\
     \yb^{\beta\,\ep}\ (1-\xb\,y)^{-1}=
   \frac{\Gamma(1+\alpha\,\ep)\Gamma(1+\beta\,\ep)}
      {\Gamma(1+(\alpha+\beta)\ep)}\ 
    \hypgeo{1}{\alpha\,\ep}{1+(\alpha+\beta)\,\ep}{\xb}\\
M_2(\alpha,\beta) =\alpha\,\ep\,&\int_0^1\ dy\,y^{-1+\alpha\,\ep}\
     \yb^{\beta\,\ep}\ \hypgeo{1}{-\ep}{1-\ep}{-\frac{y}{\yb}} =\\
  & \frac{\Gamma(1+\alpha\,\ep)\Gamma(1+(\beta-1)\,\ep)}
      {\Gamma(1+(\alpha+\beta-1)\ep)}\ 
  \ghypgeo{-\ep}{-\ep}{\alpha\,\ep}{1-\ep}{1+(\alpha+\beta-1)\,\ep}{1}\\
M_3(\alpha,\beta) =\alpha\,\ep\,&\int_0^1\ dy\,y^{-1+\alpha\,\ep}\
     \yb^{\beta\,\ep}\ \hypgeo{1}{-\ep}{1-\ep}{-x\frac{y}{\yb}} =\\
  &\frac{\Gamma(1+\alpha\,\ep)\Gamma(1+\beta\,\ep)}
      {\Gamma(1+(\alpha+\beta)\ep)}\ 
    \ghypgeo{1}{-\ep}{\alpha\,\ep}{1-\ep}{-\beta\,\ep}{x} \\
  & + \frac{\alpha\,\Gamma(1+\beta\,\ep)\Gamma(1-\beta\,\ep)}{\beta\,(\alpha+\beta)}\
      x^{\beta\,\ep}\Lx\hypgeo{(\alpha+\beta)\,\ep}{(\beta-1)\,\ep}{1+(\beta-1)\,\ep}{x}
      - \xb^{-(\alpha+\beta)\,\ep}\Rx\\
M_4(n,\alpha,\beta) = \alpha\,\ep\,&\int_0^1\ dy\,y^{-1+\alpha\,\ep}\
     \yb^{\beta\,\ep}\ \hypgeo{1}{n\,\ep}{1+n\,\ep}{\xb\,y} =\\
  &\frac{\Gamma(1+\alpha\,\ep)\Gamma(1+\beta\,\ep)}
      {\Gamma(1+(\alpha+\beta)\ep)}\ 
    \ghypgeo{1}{n\,\ep}{\alpha\,\ep}{1+n\,\ep}{1+(\alpha+\beta)\,\ep}{\xb}\\
M_5(\alpha) = \alpha\,\ep\,&\int_0^1\ dy\,y^{-1+\alpha\,\ep}\
     \yb^{\alpha\,\ep}\ \hypgeo{1}{\ep}{1+\ep}{-\frac{\xb^2\,y\,\yb}{x}} =\\
  &\frac{\Gamma^2(1+\alpha\,\ep)}{\Gamma(1+2\,\alpha\,\ep)}
    \ghypgeo{1}{\ep}{\alpha\,\ep}{\frac{1}{2}+\alpha\,\ep}{1+\ep}{-\frac{\xb^2}{4\,x}}
\end{split}
\end{equation}
It might appear that master integral $M_3$ contains factors of
$\ln(\xb\!)$.  It turns out, however, that when the hypergeometric
functions are expanded in $\ep$, the $\ln(\xb\!)$ terms contained in
the hypergeometrics exactly cancel the explicit logs from the
$\xb^{-(\alpha+\beta)\,\ep}$ terms.  Note also that the $\ep$
expansion of $M_5(\alpha)$ involves expanding around a half-integer
parameter in the hypergeometric function.  Such expansions are
discussed in Ref.~\cite{Huber:2007dx}.

\subsection{Master Integrals at \nnnlo}
There are more than twenty new master integrals that appear at \nnnlo.
A few of them, particularly those that involve the products of
hypergeometric functions of the same argument, can be computed in
closed form, although the resulting functions are still hard to expand
in $\ep$, even for tools like
\HypExp~\cite{Huber:2005yg,Huber:2007dx}.  As an example,
\begin{equation}
\begin{split}
\label{eqn:mi26}
\int_0^1\,dy\ &y^{-1-\ep}\,\yb^{-3\,\ep}
    \hypgeo{1}{-\ep}{1-\ep}{\xb\,y}\hypgeo{1}{-\ep}{1-\ep}{\xb\,y}\\
  = &  \frac{\Gamma(1-\ep)\Gamma(1-3\,\ep)}{\Gamma(1-4\,\ep)(\ep)}
         \LB1-\ghypgeo{1}{-\ep}{-\ep}{1-\ep}{1-4\,\ep}{\xb}
         \vphantom{\frac{2}{\delta}}\right.\\
    & - \lim_{\delta\to0}\frac{2\,\ep^2\,\xb\!}{\delta\,(1-2\ep)(1-4\ep)}
       \Lx\vphantom{\frac{\Gamma}{\Gamma}}
         \ghypgeo{1}{1-2\ep}{1-\ep}{2-2\ep}{2-4\ep}{\xb\!}\right.\\
    &\left.\left.\vphantom{\frac{\Gamma}{\Gamma}}\hskip110pt
       - \ghypgeo{1}{1-2\ep}{1-\ep+\delta\ep}{2-2\ep}{2-4\ep}{\xb\!}
     \Rx\RB\\
\end{split}
\end{equation}
Both for this reason, and the fact that many of the masters cannot be
evaluated in closed form, I choose to compute all of the needed integrals
directly as a Laurent series in $\ep$ by means of threshold expansion.
The exceptions are the two scale-free master integrals, $M_{10}$ and
$M_{11}$, which integrate to pure numbers,

The full list of master integrals needed for the \nnnlo\ contribution
is given below.  The coefficients are chosen so that each of the
master integrals is a function of uniform transcendentality ${\cal T}
= 0$, with the leading term in the $\ep$ expansion equal to unity.

\begin{equation}
\begin{split}
\label{eqn:MINNNLO}
M_{10} &= -\ep\int_0^1\,dy\ y^{-1-\ep}\,\yb^{-\ep}\ 
    \hypgeo{1}{-\ep}{1-\ep}{-\frac{y}{\yb}}^2\,,\\
M_{11} &= -2\,\ep\int_0^1\,dy\ y^{-1-\ep}\,\yb^{-\ep}\ 
    \hypgeo{1}{-\ep}{1-\ep}{-\frac{y}{\yb}}
    \hypgeo{1}{-\ep}{1-\ep}{-\frac{\yb}{y}}\,.\\
M_{12} &= -\ep\int_0^1\,dy\ y^{-1-\ep}\,\yb^{-\ep}\ 
   \hypgeo{1}{-\ep}{1-\ep}{-\frac{y}{\yb}}
   \hypgeo{1}{-\ep}{1-\ep}{-x\frac{y}{\yb}}\\
M_{13} &= -2\ep\int_0^1\,dy\ y^{-1-\ep}\,\yb^{-\ep}\ 
   \hypgeo{1}{-\ep}{1-\ep}{-\frac{y}{\yb}}
   \hypgeo{1}{-\ep}{1-\ep}{-x\frac{\yb}{y}}\\
M_{14}(n) &= -\ep\int_0^1\,dy\ y^{-1-\ep}\,\yb^{n\,\ep}\ 
   \hypgeo{1}{-\ep}{1-\ep}{-\frac{y}{\yb}}
   \Lx1-\xb\,y\Rx^{-1}\\
M_{15}(n) &= -2\ep\int_0^1\,dy\ y^{n\,\ep}\,\yb^{-1-\ep}\ 
   \hypgeo{1}{-\ep}{1-\ep}{-\frac{y}{\yb}}
   \Lx1-\xb\,\yb\Rx^{-1}\\
M_{16}(m) &= -\ep\int_0^1\,dy\ y^{-1-\ep}\,\yb^{-2\,\ep}\ 
   \hypgeo{1}{-\ep}{1-\ep}{-\frac{y}{\yb}}
   \hypgeo{1}{m\,\ep}{1+m\,\ep}{\xb\,y}\\
M_{17}(m) &= -2\ep\int_0^1\,dy\ y^{-1-2\ep}\,\yb^{-\ep}\ 
   \hypgeo{1}{-\ep}{1-\ep}{-\frac{y}{\yb}}
   \hypgeo{1}{m\,\ep}{1+m\,\ep}{\xb\,\yb}\\
M_{18} &= -\ep\int_0^1\,dy\ y^{-1-\ep}\,\yb^{-\ep}\ 
   \hypgeo{1}{-\ep}{1-\ep}{-\frac{y}{\yb}}
   \hypgeo{1}{\ep}{1+\ep}{-\frac{\xb^2\,y\,\yb}{x}}\\
M_{19} &= -\ep\int_0^1\,dy\ y^{-1-\ep}\,\yb^{-\ep}\ 
   \hypgeo{1}{-\ep}{1-\ep}{-x\frac{y}{\yb}}^2\\
M_{20} &= -2\ep\int_0^1\,dy\ y^{-1-\ep}\,\yb^{-\ep}\ 
   \hypgeo{1}{-\ep}{1-\ep}{-x\frac{y}{\yb}}
   \hypgeo{1}{-\ep}{1-\ep}{-x\frac{\yb}{y}}\\
\end{split}
\end{equation}
\begin{equation*}
\begin{split}
M_{21}(n) &= -\ep\int_0^1\,dy\ y^{-1-\ep}\,\yb^{n\,\ep}\ 
   \hypgeo{1}{-\ep}{1-\ep}{-x\frac{y}{\yb}}
   \Lx1-\xb\,y\Rx^{-1}\\
M_{22}(n) &= -2\ep\int_0^1\,dy\ y^{n\,\ep}\,\yb^{-1-\ep}\ 
   \hypgeo{1}{-\ep}{1-\ep}{-x\frac{y}{\yb}}
   \Lx1-\xb\,\yb\Rx^{-1}\\
M_{23}(m) &= -\ep\int_0^1\,dy\ y^{-1-\ep}\,\yb^{-2\,\ep}\ 
   \hypgeo{1}{-\ep}{1-\ep}{-x\frac{y}{\yb}}
   \hypgeo{1}{m\,\ep}{1+m\,\ep}{\xb\,y}\\
M_{24}(m) &= -2\ep\int_0^1\,dy\ y^{-1-2\ep}\,\yb^{-\ep}\ 
   \hypgeo{1}{-\ep}{1-\ep}{-x\frac{y}{\yb}}
   \hypgeo{1}{m\,\ep}{1+m\,\ep}{\xb\,\yb}\\
M_{25} &= -\ep\int_0^1\,dy\ y^{-1-\ep}\,\yb^{-\ep}\ 
   \hypgeo{1}{-\ep}{1-\ep}{-x\frac{y}{\yb}}
   \hypgeo{1}{\ep}{1+\ep}{-\frac{\xb^2\,y\,\yb}{x}}\\
M_{26}(n,m) &= -\ep\int_0^1\,dy\ y^{-1-\ep}\,\yb^{-3\ep}\ 
   \hypgeo{1}{n\,\ep}{1+n\,\ep}{\xb\,y}
   \hypgeo{1}{m\,\ep}{1+m\,\ep}{\xb\,y}\\
M_{27}(n,m) &= -2\ep\int_0^1\,dy\ y^{-1-2\ep}\,\yb^{-2\ep}\ 
   \hypgeo{1}{n\,\ep}{1+n\,\ep}{\xb\,y}
   \hypgeo{1}{m\,\ep}{1+m\,\ep}{\xb\,\yb}\\
M_{28}(n,m) &= -\ep\int_0^1\,dy\ y^{-1-\ep}\,\yb^{n\,\ep}\ 
   \hypgeo{1}{m\,\ep}{1+m\,\ep}{\xb\,y}
   \Lx1-\xb\,y\Rx^{-1}\\
M_{29}(n,m) &= -2\ep\int_0^1\,dy\ y^{n\,\ep}\,\yb^{-1-2\ep}\ 
   \hypgeo{1}{m\,\ep}{1+m\,\ep}{\xb\,y}
   \Lx1-\xb\,\yb\Rx^{-1}\\
M_{30}(n) &= -\ep\int_0^1\,dy\ y^{-1-\ep}\,\yb^{-2\,\ep}\ 
   \hypgeo{1}{n\,\ep}{1+n\,\ep}{\xb\,y}
   \hypgeo{1}{\ep}{1+\ep}{-\frac{\xb^2\,y\,\yb}{x}}\\
M_{31} &= -\ep\int_0^1\,dy\ y^{-1-\ep}\,\yb^{-\ep}\ 
   \hypgeo{1}{\ep}{1+\ep}{-\frac{\xb^2\,y\,\yb}{x}}^2\\
M_{32}(n) &= -\ep\int_0^1\,dy\ y^{-1-\ep}\,\yb^{n\,\ep}\ 
   \hypgeo{1}{\ep}{1+\ep}{-\frac{\xb^2\,y\,\yb}{x}}
   \Lx1-\xb\,y\Rx^{-1}
\end{split}
\end{equation*}

In addition, one also needs a variation on $M_5$,
\begin{equation}
M_{6}(\alpha,\beta) = \alpha\,\ep\,\int_0^1\ dy\,y^{-1+\alpha\,\ep}\
     \yb^{\beta\,\ep}\ \hypgeo{1}{\ep}{1+\ep}{-\frac{\xb^2\,y\,\yb}{x}}\,,
\end{equation}
where $M_5(\alpha) = M_6(\alpha,\alpha)$.  Note that while $M_5$ can be
expressed in closed form, $M_6$ cannot.

\subsection{Threshold expansions of the integrands}
The threshold expansion of the integrands is quite simple.  In many
cases, one can simply use the series representation of the
hypergeometric function
\begin{equation}
  \hypgeo{\alpha}{\beta}{\gamma}{z} = \sum_{n=0}^\infty
   \frac{(a)_n\,(b)_n}{n!\,(c)_n}\,z^n\,,
\end{equation}
where $(a)_n$ is the Pochhammer symbol
\begin{equation}
(a)_n = \frac{\Gamma(a+n)}{\Gamma(a)}\,.
\end{equation}
This works well for hypergeometric functions of argument $(\xb\,y)$ and
$(\xb\,\yb)$.  It also works for the hypergeometrics of argument $(-
x^{-1}\,\xb^2\,y\,\yb)$ if one then expands the resulting factors of
$x^{-m}$,
\begin{equation}
x^{-m} = (1-\xb)^{-m} = \hypgeo{m}{a}{a}{\xb} = \sum_{n=0}^\infty
   \frac{(m)_n}{n!}\,\xb^n\,.
\end{equation}
In the same way, factors of $(1-\xb\,y)^{-m}$ are expanded as
\begin{equation}
\Lx1-\xb\,y\Rx^{-m} = \hypgeo{m}{a}{a}{\xb\,y} = \sum_{n=0}^\infty
   \frac{(m)_n}{n!}\,\Lx\xb\,y\Rx^n\,.
\end{equation}
The only terms that don't expand trivially in this way are the
hypergeometrics with arguments $\Lx-x\,y/\yb\Rx$ and
$\Lx-x\,\yb/y\Rx$.  For these, one simply uses the Taylor series
expansion,
\begin{equation}
  \hypgeo{1}{-\ep}{1-\ep}{-x\frac{y}{\yb}} = \sum_{n=0}^\infty
    \frac{\xb^n}{n!}\ \LB\frac{d^n}{d\xb^n}
     \hypgeo{1}{-\ep}{1-\ep}{(\xb-1)\frac{y}{\yb}}\RB_{\xb=0}\,,
\end{equation}
where
\begin{equation}
  \frac{d}{d\xb}\hypgeo{a}{b}{c}{{(\xb-1)\frac{y}{\yb}}} =
     \frac{y}{\yb}\frac{a\,b}{c}
     \hypgeo{a+1}{b+1}{c+1}{{(\xb-1)\frac{y}{\yb}}}\,.
\end{equation}
Combining these equations and repeatedly applying hypergeometric
identities for contiguous functions (see, {\it e.g.}\
Ref.~\cite{Gradshteyn:Tables}), I obtain the threshold expansion to be
\begin{equation}
  \hypgeo{1}{-\ep}{1-\ep}{-x\frac{y}{\yb}} = \sum_{n=0}^\infty
    \frac{\xb^n}{n!}\ \frac{(-\ep)_n}{n!}\Lx
    \hypgeo{1}{-\ep}{1-\ep}{-\frac{y}{\yb}}
       - \yb\,\sum_{m=0}^{n-1}\ y^m\,\frac{m!}{(1-\ep)_m}\Rx\,.
\end{equation}

Thus, when the threshold expansion is performed on all components of
the integrands, the result is a sum of powers of $\xb$ multiplying
integrals in $y$ and $\yb$ only.  These integrals can all be reduced to
combinations of master integrals $M_0$, $M_2$, $M_{10}$ and $M_{11}$,
given in Eqs.~(\ref{eqn:MINLO}),\,(\ref{eqn:MINNLO}) and
(\ref{eqn:MINNNLO}).

\subsection{Results for the Partonic Cross Sections}
The results of these calculations are merely parts of a physical
result, namely the inclusive Higgs production cross section to
\nnnlo.  By themselves, they have no direct physical interpretation.
Thus, while I have described how one would perform $\msbar$
renormalization on these terms, I present the results of the bare
calculation, and leave renormalization until such time as all pieces
of the \nnnlo\ cross section can be assembled.

The contributions can be broken into two distinct components, the soft
and the hard contributions.  The soft contributions come entirely from
the leading behavior in $\xb$, that is terms that go like
$\xb^{-1+n\,\ep}$, which can be expanded in distributions as described
\Sec{sec:SRE}.  The hard contribution is comprised of all other terms.
Only the purely gluon-initiated partonic cross section $g\,g\to\ H\,g$,
has soft contributions.

\subsubsection{Contributions starting at \nlo}
The contribution to the inclusive cross section from the square of
tree-level amplitudes starts at \nlo\ and, through the renormalization
of $\alpha_s$, the effective operator ${\cal O}_1$ and the Wilson
coefficient $C_1$, applies to all higher orders.  The results of this
calculation depend only on master integral $M_0$, which expands
readily to arbitrary order in $\ep$.
\begin{equation}
\begin{split}
  \sigma_{gg\to H\,g}^{1,B} =\ & \Cfact\,\abare\mumh^{\ep}\,M_{0}(-1,-1)\LB
            \frac{3\,\delta(\xb)}{\ep^2\,(1-\ep)}
          - \frac{6\,\DSum{-2}\,x^{\ep}}{\ep\,(1-\ep)}\right.\\
    + &\left.  \frac{x^{\ep}\,\xb^{-2\,\ep}}{\ep}\Lx
          \frac{12}{1-\ep}
          - \xb\,\frac{18 - 54\,\ep + 42\,\ep^2}{(1-\ep)^2\,(1-2\,\ep)}
          + \xb^2\,\frac{12 - 36\,\ep + 30\,\ep^2}{(1-\ep)^2\,(1-2\,\ep)}
          - \xb^3\,\frac{36 - 27\,\ep}{2\,(1-2\,\ep)\,(3-2\,\ep)}\Rx\RB\\
   \sigma_{q\overline{q}\to H\,g}^{1,B} =\ & \Cfact\,\abare\mumh^{\ep}\,M_{0}(-1,-1)\,
       x^{\ep}\,\xb^{-2\,\ep}\,\xb^3\,\frac{32\,(1-\ep)^2}{9\,(1-2\,\ep)\,(3-2\,\ep)}\\
   \sigma_{gq\to H\,q}^{1,B} =\ & -\Cfact\,\abare\mumh^{\ep}\,M_{0}(-1,-1)\,
       x^{\ep}\,\xb^{-2\,\ep}\,\Lx\frac{2}{3\,\ep} +
       \xb\,\frac{4}{3\,(1-2\,\ep)} + \xb^2\frac{2-\ep}{3\,\ep\,(1-2\,\ep)}\Rx      
\end{split}
\end{equation}
where, as in \eqn{eqn:xbdistrib}, ${\cal D}^{-2\,\ep}(\xb)$ represents
the tower of plus-distributions in $\xb$ weighted by $(-2\,\ep)$.
Using the expansion of $M_{0}(-1,-1)$ given in \eqn{eqn:M0exp}, one
easily recovers the previously known results for these terms.

\subsubsection{Contributions starting at \nnlo}
The contribution from the interference of tree-level and one-loop
amplitudes starts at \nnlo\ and, through renormalization, contributes
to all higher orders.  The results of this calculation depend on six
master integrals, $M_{0-5}$, which are all known in closed form (see
\eqn{eqn:MINNLO}).  In addition to the phase space integrals, there
are products of $\Gamma$-functions that arise from the loop
integration that can be cast into the same form as the master integral
$M_0(\alpha,\beta)$.
{\small
\begin{equation*}
\begin{split}
  \sigma_{gg\to H\,g}^{2,B} =\ &
     \Cfact\,\abare^2\mumh^{2\,\ep}\left\{ \vphantom{\frac{9}{\ep^2}}\right.\\
  &   \frac{\Gamma^5(1-\ep)\,\Gamma^3(1+\ep)}{\Gamma^2(1-2\,\ep)\,\Gamma(1+2\,\ep)}
      \,M_{0}(-2,-2)\Lx
          - \frac{9\,\delta(\xb)}{8\,\ep^4\,(1-\ep)}
          + \frac{9\,\DSum{-4}\,x^{2\,\ep}}{2\,\ep^3\,(1-\ep)}
          + x^{2\,\ep}\,\xb^{-4\,\ep}\Lx
          - \frac{9}{\ep^3\,(1-\ep)}\right.\right.\\
  &\qquad \left.\left.
          + \xb\,\frac{27 - 135\,\ep + 135\,\ep^2}{2\,\ep^3\,(1-\ep)^2\,(1-4\,\ep)}
          - \xb^2\,\frac{18 - 90\,\ep + 99\,\ep^2}{2\,\ep^3\,(1-\ep)^2\,(1-4\,\ep)}
          + \xb^3\,\frac{54 - 189\,\ep + 162\,\ep^2}{4\,\ep^3\,(1-\ep)\,(1-4\,\ep)\,(3-4\,\ep)}
           \Rx\Rx\\
  +& \frac{\Gamma^4(1-\ep)\,\Gamma^2(1+\ep)}{\Gamma^2(1-2\,\ep)\,\Gamma(1+2\,\ep)}\LB
       M_{0}(-1,-1)\Lx
          - \delta(\xb)\,\frac{9 - 27\,\ep + 18\,\ep^2 +
            9\,\ep^3}{\ep^4\,(1-\ep)^2\,(1-2\,\ep)}
\right.\right.\\&\qquad
          + \DSum{-2}\,x^{\ep}\,\frac{18 - 108\,\ep + 234\,\ep^2 - 198\,\ep^3 + 27\,\ep^4 + 36\,\ep^5}{\ep^3\,(1-\ep)^3\,(1-2\,\ep)^2}
          - \DSum{-2}\,x^{2\,\ep}\,\frac{9 - 54\,\ep + 117\,\ep^2 - 108\,\ep^3 + 54\,\ep^4}{2\,\ep^3\,(1-\ep)^3\,(1-2\,\ep)^2}
\\&\qquad
        + x^{\ep}\,\xb^{-2\,\ep}\Lx
          - \frac{324 - 3834\,\ep + 18810\,\ep^2 - 50400\,\ep^3 + 79650\,\ep^4 - 72387\,\ep^5 + 31050\,\ep^6 - 432\,\ep^7 - 2592\,\ep^8}{\ep^3\,(1-\ep)^3\,(1-2\,\ep)^2\,(1-4\,\ep)\,(3-2\,\ep)\,(3-4\,\ep)}
\right.\\&\qquad
          + \xb\,\frac{324 - 3402\,\ep + 14247\,\ep^2 - 30618\,\ep^3 + 32562\,\ep^4 - 9243\,\ep^5 - 10080\,\ep^6 + 5616\,\ep^7}{2\,\ep^3\,(1-\ep)^3\,(1-2\,\ep)^2\,(1-4\,\ep)\,(3-2\,\ep)}
\\&\qquad
          - \xb^2\,\frac{648 - 7128\,\ep + 32166\,\ep^2 - 77544\,\ep^3 + 102726\,\ep^4 - 60363\,\ep^5 - 12411\,\ep^6 + 33840\,\ep^7 - 11664\,\ep^8}{2\,\ep^3\,(1-\ep)^3\,(1-2\,\ep)^2\,(1-4\,\ep)\,(3-2\,\ep)\,(3-4\,\ep)}
\\&\qquad\left.
          + \xb^3\,\frac{108 - 405\,\ep + 459\,\ep^2 - 54\,\ep^3 - 81\,\ep^4}{2\,\ep^3\,(1-\ep)\,(1-2\,\ep)^2\,(3-2\,\ep)}
           \Rx\\&\qquad
        + x^{2\,\ep}\,\xb^{-2\,\ep}\Lx
          \frac{9 - 54\,\ep + 117\,\ep^2 - 108\,\ep^3 + 54\,\ep^4}{\ep^3\,(1-\ep)^3\,(1-2\,\ep)^2}
          - \xb\,\frac{81 - 540\,\ep + 1323\,\ep^2 - 1602\,\ep^3 + 1026\,\ep^4 - 270\,\ep^5}{2\,\ep^3\,(1-\ep)^3\,(1-2\,\ep)^2\,(3-2\,\ep)}
\right.\\&\qquad\left.
          + \xb^2\,\frac{27 - 153\,\ep + 279\,\ep^2 - 243\,\ep^3 + 81\,\ep^4}{\ep^3\,(1-\ep)^2\,(1-2\,\ep)^2\,(3-2\,\ep)}
          - \xb^3\,\frac{162 - 1323\,\ep + 3123\,\ep^2 - 3276\,\ep^3 + 1593\,\ep^4 - 297\,\ep^5}{4\,\ep^3\,(1-\ep)^2\,(1-2\,\ep)^2\,(3-2\,\ep)^2}
           \Rx\\&\qquad\left.
        + x^{2\,\ep}\,\xb^{-2\,\ep}\,N_{f}\,\Lx
          - \xb\,\frac{3}{2\,(1-\ep)^2\,(1-2\,\ep)^2\,(3-2\,\ep)}
          + \xb^2\,\frac{3}{2\,(1-\ep)^2\,(1-2\,\ep)^2\,(3-2\,\ep)}
\right.\right.\\&\qquad\left.\left.
          - \xb^3\,\frac{9 - 9\,\ep + 3\,\ep^2}{4\,(1-\ep)^3\,(1-2\,\ep)^2\,(3-2\,\ep)^2}
           \Rx\Rx
\\&
       + M_{1}(-1,-1)\Lx
          - \DSum{-2}\,x^{\ep}\,\frac{9}{2\,\ep^3\,(1-\ep)}
\right.\\&\qquad
        + x^{\ep}\,\xb^{-2\,\ep}\Lx
            \frac{81 - 756\,\ep + 3231\,\ep^2 - 8568\,\ep^3 + 13536\,\ep^4 - 10800\,\ep^5 + 3168\,\ep^6}{\ep^3\,(1-\ep)^3\,(1-2\,\ep)\,(1-4\,\ep)\,(3-2\,\ep)\,(3-4\,\ep)}
\right.\\&\qquad
          - \xb\,\frac{243 - 2268\,\ep + 9747\,\ep^2 - 24750\,\ep^3 + 31464\,\ep^4 - 9504\,\ep^5 - 12384\,\ep^6 + 6912\,\ep^7}{2\,\ep^3\,(1-\ep)^3\,(1-2\,\ep)\,(1-4\,\ep)\,(3-2\,\ep)\,(3-4\,\ep)}
\\&\qquad
          + \xb^2\,\frac{162 - 972\,\ep + 2223\,\ep^2 - 3519\,\ep^3 + 1710\,\ep^4 + 8820\,\ep^5 - 14616\,\ep^6 + 5760\,\ep^7}{2\,\ep^3\,(1-\ep)^3\,(1-2\,\ep)\,(1-4\,\ep)\,(3-2\,\ep)\,(3-4\,\ep)}
\\&\qquad\left.\left.
          - \xb^3\,\frac{162 - 729\,\ep - 360\,\ep^2 + 5445\,\ep^3 - 9954\,\ep^4 + 10332\,\ep^5 - 7416\,\ep^6 + 2304\,\ep^7}{4\,\ep^3\,(1-\ep)^3\,(1-2\,\ep)\,(1-4\,\ep)\,(3-2\,\ep)\,(3-4\,\ep)}
           \Rx\Rx
\\&
       +\Lx M_{2}(-1,-1)\,x^{2\,\ep} - M_{3}(-1,-1)\,x^{\ep}\Rx
            \frac{9\,\DSum{-2} - \xb^{-2\,\ep}\Lx
               18 - 27\,\xb + 18\,\xb^2 - 9\,\xb^3\Rx}{\ep^3\,(1-\ep)}
\\&
       + M_{5}(-1)\Lx
            \DSum{-2}\,x^{\ep}\,\frac{9}{\ep^3\,(1-\ep)}
        + x^{\ep}\,\xb^{-2\,\ep}\Lx
          - \frac{18}{\ep^3\,(1-\ep)}
          + \xb\,\frac{27 - 135\,\ep + 135\,\ep^2}{\ep^3\,(1-\ep)^2\,(1-4\,\ep)}
\right.\right.\\&\qquad\left.\left.\left.
          - \xb^2\,\frac{18 - 90\,\ep + 99\,\ep^2}{\ep^3\,(1-\ep)^2\,(1-4\,\ep)}
          + \xb^3\,\frac{54 - 189\,\ep + 162\,\ep^2}{2\,\ep^3\,(1-\ep)\,(1-4\,\ep)\,(3-4\,\ep)}
           \Rx\Rx
       \RB
\end{split}
\end{equation*}
\begin{equation}
\begin{split}
\phantom{\sigma_{gg\to H\,g}^{2,B} =\ }
+&\frac{\Gamma^3(1-\ep)\,\Gamma(1+\ep)}{\Gamma(1-2\,\ep)}\LB
         M_{0}(-1,-2)\,x^{2\,\ep}\,\xb^{-3\,\ep}\Lx
            \frac{216 + 675\,\ep - 11403\,\ep^2 + 31536\,\ep^3 - 31824\,\ep^4 + 10368\,\ep^5}{4\,\ep^2\,(1-\ep)^3\,(1-2\,\ep)\,(1-4\,\ep)\,(3-2\,\ep)\,(3-4\,\ep)}
\right.\right.\\&\qquad
          - \xb\,\frac{162 - 378\,\ep - 4347\,\ep^2 + 17136\,\ep^3 - 7641\,\ep^4 - 41220\,\ep^5 + 57888\,\ep^6 - 20736\,\ep^7}{2\,\ep^2\,(1-\ep)^3\,(1-2\,\ep)\,(1-3\,\ep)\,(1-4\,\ep)\,(3-2\,\ep)\,(3-4\,\ep)}
\\&\qquad\left.
          + \xb^2\,\frac{2592 - 29106\,\ep + 134829\,\ep^2 - 344790\,\ep^3 + 559035\,\ep^4 - 635688\,\ep^5 + 521784\,\ep^6 - 271728\,\ep^7 + 62208\,\ep^8}{4\,\ep^2\,(1-\ep)^3\,(1-2\,\ep)\,(1-3\,\ep)\,(1-4\,\ep)\,(2-3\,\ep)\,(3-2\,\ep)\,(3-4\,\ep)}
           \Rx
\\&
       + M_{0}(-1,-2)\,x^{2\,\ep}\,\xb^{-3\,\ep}\,N_{f}\,\Lx
          - \frac{3}{4\,\ep\,(1-\ep)^3\,(1-2\,\ep)\,(3-2\,\ep)}
          + \xb\,\frac{3}{2\,\ep\,(1-\ep)^2\,(1-2\,\ep)\,(3-2\,\ep)}
\right.\\&\qquad\left.
          - \xb^2\,\frac{6 - 27\,\ep + 36\,\ep^2}{4\,\ep\,(1-\ep)^2\,(1-2\,\ep)\,(1-3\,\ep)\,(2-3\,\ep)\,(3-2\,\ep)}
           \Rx
\\&
       + M_{1}(-1,-2) \Lx
            \frac{9\,\DSum{-3}\,x^{2\,\ep}}{\ep^3\,(1-\ep)}
        + x^{2\,\ep}\,\xb^{-3\,\ep}\Lx
          - \frac{162 - 1566\,\ep + 6300\,\ep^2 - 14328\,\ep^3 + 19260\,\ep^4 - 13680\,\ep^5 + 3744\,\ep^6}{\ep^3\,(1-\ep)^3\,(1-2\,\ep)\,(1-4\,\ep)\,(3-2\,\ep)\,(3-4\,\ep)}
\right.\right.\\&\qquad
          + \xb\,\frac{486 - 4698\,\ep + 19197\,\ep^2 - 43245\,\ep^3 + 50796\,\ep^4 - 19764\,\ep^5 - 10224\,\ep^6 + 6912\,\ep^7}{2\,\ep^3\,(1-\ep)^3\,(1-2\,\ep)\,(1-4\,\ep)\,(3-2\,\ep)\,(3-4\,\ep)}
\\&\qquad
          - \xb^2\,\frac{162 - 1296\,\ep + 4302\,\ep^2 - 8127\,\ep^3 + 7659\,\ep^4 + 720\,\ep^5 - 6516\,\ep^6 + 2880\,\ep^7}{\ep^3\,(1-\ep)^3\,(1-2\,\ep)\,(1-4\,\ep)\,(3-2\,\ep)\,(3-4\,\ep)}
\\&\qquad\left.\left.
          + \xb^3\,\frac{162 - 1026\,\ep + 2007\,\ep^2 - 1017\,\ep^3 - 1494\,\ep^4 + 3492\,\ep^5 - 3384\,\ep^6 + 1152\,\ep^7}{2\,\ep^3\,(1-\ep)^3\,(1-2\,\ep)\,(1-4\,\ep)\,(3-2\,\ep)\,(3-4\,\ep)}
           \Rx\Rx
\\&
       + M_{4}(-1,-1,-2)\,x^{2\,\ep}
            \frac{9\,\DSum{-3} - \xb^{-3\,\ep}\Lx18 - 27\,\xb + 18\,\xb^2-9\,\xb^3\Rx}{\ep^3\,(1-\ep)}
\\&
       + M_{4}(1,-1,-2) \Lx
          - \frac{18\,\DSum{-3}\,x^{2\,\ep}}{\ep^3\,(1-\ep)}
        + x^{2\,\ep}\,\xb^{-3\,\ep}\Lx
          + \frac{36}{\ep^3\,(1-\ep)}
          - \xb\,\frac{54 - 270\,\ep + 270\,\ep^2}{\ep^3\,(1-\ep)^2\,(1-4\,\ep)}
\right.\right.\\&\qquad\left.\left.\left.\left.
          + \xb^2\,\frac{36 - 180\,\ep + 198\,\ep^2}{\ep^3\,(1-\ep)^2\,(1-4\,\ep)}
          - \xb^3\,\frac{54 - 189\,\ep + 162\,\ep^2}{\ep^3\,(1-\ep)\,(1-4\,\ep)\,(3-4\,\ep)}
           \Rx\Rx
\RB\right\}
\end{split}
\end{equation}
\begin{equation}
\begin{split}
  \sigma_{q\overline{q}\to H\,g}^{2,B} =\ & \Cfact\,\abare^2\mumh^{2\ep}\left\{
       \vphantom{\frac{9}{\ep^2}}\right.\\&
   \frac{\Gamma^5(1-\ep)\,\Gamma^3(1+\ep)}{\Gamma^2(1-2\,\ep)\,\Gamma(1+2\,\ep)}\LB
        M_{0}(-2,-2)\,x^{2\,\ep}\,\xb^{-4\,\ep}\Lx
            \xb^2\,\frac{8}{27\,(1-4\,\ep)}
          + \xb^3\,\frac{16 - 40\,\ep + 8\,\ep^2 + 32\,\ep^3}{27\,\ep^2\,(1-4\,\ep)\,(3-4\,\ep)}
          \Rx\RB
\\+&
   \frac{\Gamma^4(1-\ep)\,\Gamma^2(1+\ep)}{\Gamma^2(1-2\,\ep)\,\Gamma(1+2\,\ep)}\LB
        M_{0}(-1,-1)\,x^{\ep}\,\xb^{-2\,\ep}\Lx
            \frac{768 - 176\,\ep - 2720\,\ep^2 + 2176\,\ep^3}{27\,\ep^2\,(1-2\,\ep)\,(3-2\,\ep)\,(3-4\,\ep)}
\right.\right.\\&\qquad
          - \xb\,\frac{384 - 2128\,\ep + 1872\,\ep^2 + 2608\,\ep^3 - 1856\,\ep^4 - 2624\,\ep^5 + 2176\,\ep^6}{27\,\ep^2\,(1-\ep)\,(1-2\,\ep)^2\,(1-4\,\ep)\,(3-2\,\ep)}
\\&\qquad
          + \xb^2\,\frac{1152 - 10512\,\ep + 37648\,\ep^2 - 77264\,\ep^3 + 100352\,\ep^4 - 77024\,\ep^5 + 29312\,\ep^6 - 4096\,\ep^7}{27\,\ep^2\,(1-\ep)\,(1-2\,\ep)^2\,(1-4\,\ep)\,(3-2\,\ep)\,(3-4\,\ep)}
\\&\qquad\left.
          - \xb^3\,\frac{32 - 128\,\ep + 160\,\ep^2 - 32\,\ep^3 - 32\,\ep^4}{3\,\ep^2\,(1-2\,\ep)^2\,(3-2\,\ep)}
           \Rx
\\&
       + M_{0}(-1,-1)\,x^{2\,\ep}\,\xb^{-2\,\ep}\Lx
          - \xb\,\frac{16}{3\,(1-2\,\ep)^2\,(3-2\,\ep)}
          + \xb^2\,\frac{16}{3\,(1-2\,\ep)^2\,(3-2\,\ep)}
\right.\\&\qquad\left.
          + \xb^3\,\frac{48 - 200\,\ep + 160\,\ep^2 - 224\,\ep^3 + 272\,\ep^4 - 160\,\ep^5 + 32\,\ep^6}{27\,\ep^2\,(1-\ep)\,(1-2\,\ep)^2\,(3-2\,\ep)^2}
           \Rx
\\&
       + M_{0}(-1,-1)\,x^{2\,\ep}\,\xb^{-2\,\ep}\,N_{f}\,\Lx
          - \xb^3\,\frac{32 - 96\,\ep + 96\,\ep^2 - 32\,\ep^3}{9\,\ep\,(1-2\,\ep)^2\,(3-2\,\ep)^2}
           \Rx
\\&
       + M_{1}(-1,-1)\,x^{\ep}\,\xb^{-2\,\ep}\Lx
          - \frac{768 - 176\,\ep - 2720\,\ep^2 + 2176\,\ep^3}{27\,\ep^2\,(1-2\,\ep)\,(3-2\,\ep)\,(3-4\,\ep)}
\right.\\&\qquad
          + \xb\,\frac{1152 - 6384\,\ep + 5808\,\ep^2 + 7984\,\ep^3 - 10496\,\ep^4 - 1984\,\ep^5 + 4352\,\ep^6}{27\,\ep^2\,(1-\ep)\,(1-2\,\ep)\,(1-4\,\ep)\,(3-2\,\ep)\,(3-4\,\ep)}
\\&\qquad
          - \xb^2\,\frac{1152 - 8976\,\ep + 24648\,\ep^2 - 36472\,\ep^3 + 31552\,\ep^4 - 15520\,\ep^5 + 4480\,\ep^6}{27\,\ep^2\,(1-\ep)\,(1-2\,\ep)\,(1-4\,\ep)\,(3-2\,\ep)\,(3-4\,\ep)}
\\&\qquad\left.
          + \xb^3\,\frac{816 - 6904\,\ep + 20800\,\ep^2 - 30280\,\ep^3 + 20832\,\ep^4 - 4960\,\ep^5 + 128\,\ep^6}{27\,\ep^2\,(1-\ep)\,(1-2\,\ep)\,(1-4\,\ep)\,(3-2\,\ep)\,(3-4\,\ep)}
           \Rx
\\&\left.
       + M_{5}(-1)\,x^{\ep}\,\xb^{-2\,\ep}\Lx
            \xb^2\,\frac{16}{27\,(1-4\,\ep)}
          + \xb^3\,\frac{32 - 80\,\ep + 16\,\ep^2 + 64\,\ep^3}{27\,\ep^2\,(1-4\,\ep)\,(3-4\,\ep)}
           \Rx\RB
\\+&
    \frac{\Gamma^3(1-\ep)\,\Gamma(1+\ep)}{\Gamma(1-2\,\ep)}\LB
         M_{0}(-1,-2)\,x^{2\,\ep}\,\xb^{-3\,\ep}\Lx
          - \frac{768 - 176\,\ep - 2720\,\ep^2 + 2176\,\ep^3}{27\,\ep^2\,(1-2\,\ep)\,(3-2\,\ep)\,(3-4\,\ep)}
\right.\right.\\&\qquad
          + \xb\,\frac{1152 - 9072\,\ep + 20944\,\ep^2 - 8208\,\ep^3 - 19376\,\ep^4 + 7744\,\ep^5 + 19008\,\ep^6 - 13056\,\ep^7}{27\,\ep^2\,(1-\ep)\,(1-2\,\ep)\,(1-3\,\ep)\,(1-4\,\ep)\,(3-2\,\ep)\,(3-4\,\ep)}
\\&\qquad\left.
          - \xb^2\,\frac{2304 - 26784\,\ep + 128000\,\ep^2 - 340896\,\ep^3 + 568912\,\ep^4 - 614432\,\ep^5 + 416224\,\ep^6 - 163520\,\ep^7 + 31488\,\ep^8}{27\,\ep^2\,(1-\ep)\,(1-2\,\ep)\,(1-3\,\ep)\,(1-4\,\ep)\,(2-3\,\ep)\,(3-2\,\ep)\,(3-4\,\ep)}
           \Rx
\\&
       + M_{1}(-1,-2)\,x^{2\,\ep}\,\xb^{-3\,\ep}\Lx
            \frac{768 - 176\,\ep - 2720\,\ep^2 + 2176\,\ep^3}{27\,\ep^2\,(1-2\,\ep)\,(3-2\,\ep)\,(3-4\,\ep)}
\right.\\&\qquad
          - \xb\,\frac{1152 - 6384\,\ep + 5808\,\ep^2 + 7984\,\ep^3 - 10496\,\ep^4 - 1984\,\ep^5 + 4352\,\ep^6}{27\,\ep^2\,(1-\ep)\,(1-2\,\ep)\,(1-4\,\ep)\,(3-2\,\ep)\,(3-4\,\ep)}
\\&\qquad
          + \xb^2\,\frac{1152 - 8976\,\ep + 24720\,\ep^2 - 36832\,\ep^3 + 32192\,\ep^4 - 16000\,\ep^5 + 4608\,\ep^6}{27\,\ep^2\,(1-\ep)\,(1-2\,\ep)\,(1-4\,\ep)\,(3-2\,\ep)\,(3-4\,\ep)}
\\&\qquad\left.
          - \xb^3\,\frac{768 - 6608\,\ep + 20144\,\ep^2 - 29744\,\ep^3 + 20928\,\ep^4 - 5312\,\ep^5 + 256\,\ep^6}{27\,\ep^2\,(1-\ep)\,(1-2\,\ep)\,(1-4\,\ep)\,(3-2\,\ep)\,(3-4\,\ep)}
           \Rx
\\&\left.\left.
       - M_{4}(1,-1,-2)\,x^{2\,\ep}\,\xb^{-3\,\ep}\Lx
            \xb^2\,\frac{32}{27\,(1-4\,\ep)}
          + \xb^3\,\frac{64 - 160\,\ep + 32\,\ep^2 + 128\,\ep^3}{27\,\ep^2\,(1-4\,\ep)\,(3-4\,\ep)}
           \Rx\RB\right\}
\end{split}
\end{equation}
\begin{equation}
\begin{split}
  \sigma_{gq\to H\,q}^{2,B} =\ & \Cfact\,\abare^2\mumh^{2\,\ep}\left\{
       \vphantom{\frac{9}{\ep^2}}\right.\\&
   \frac{\Gamma^5(1-\ep)\,\Gamma^3(1+\ep)}{\Gamma^2(1-2\,\ep)\,\Gamma(1+2\,\ep)}\LB
         M_{0}(-2,-2)\,x^{2\,\ep}\,\xb^{-4\,\ep}\Lx
            \frac{1}{2\,\ep^3}
          + \xb\,\frac{1}{\ep^2\,(1-4\,\ep)}
          + \xb^2\,\frac{1-2\,\ep - \ep^2}{2\,\ep^3\,(1-\ep)\,(1-4\,\ep)}
           \Rx\RB
\\+&
   \frac{\Gamma^4(1-\ep)\,\Gamma^2(1+\ep)}{\Gamma^2(1-2\,\ep)\,\Gamma(1+2\,\ep)}\LB
         M_{0}(-1,-1)\,x^{\ep}\,\xb^{-2\,\ep}\Lx
            \frac{8 - 85\,\ep + 235\,\ep^2 + 55\,\ep^3 - 995\,\ep^4 + 974\,\ep^5 + 24\,\ep^6}{9\,\ep^3\,(1-\ep)^2\,(1-2\,\ep)^2\,(1-4\,\ep)}
\right.\right.\\&\left.\qquad
          - \xb\,\frac{18 - 147\,\ep + 441\,\ep^2 - 322\,\ep^3 - 214\,\ep^4 + 8\,\ep^5}{9\,\ep^2\,(1-\ep)^2\,(1-2\,\ep)^2\,(1-4\,\ep)}
          + \xb^2\,\frac{8 - 28\,\ep + 20\,\ep^2 - 5\,\ep^3-4\,\ep^4}{9\,\ep^3\,(1-\ep)\,(1-2\,\ep)^2}
           \Rx
\\&
       + M_{0}(-1,-1)\,x^{2\,\ep}\,\xb^{-2\,\ep}\Lx
            \frac{11 - 55\,\ep + 86\,\ep^2 - 36\,\ep^3 - 24\,\ep^4}{18\,\ep^3\,(1-\ep)\,(1-2\,\ep)^2}
          + \xb\,\frac{11 - 24\,\ep + 26\,\ep^2 + 14\,\ep^3}{9\,\ep^2\,(1-\ep)\,(1-2\,\ep)^2}
\right.\\&\left.\qquad
          + \xb^2\,\frac{11 - 28\,\ep + 34\,\ep^2 - 16\,\ep^3 - 9\,\ep^4}{18\,\ep^3\,(1-\ep)\,(1-2\,\ep)^2}
           \Rx
\\&
       + M_{1}(-1,-1)\,x^{\ep}\,\xb^{-2\,\ep}\Lx
          - \frac{9 - 82\,\ep + 45\,\ep^2 + 630\,\ep^3 - 914\,\ep^4 - 120\,\ep^5}{18\,\ep^3\,(1-\ep)^2\,(1-2\,\ep)\,(1-4\,\ep)}
          + \xb\,\frac{25 - 159\,\ep + 286\,\ep^2 + 88\,\ep^3}{9\,\ep^2\,(1-\ep)\,(1-2\,\ep)\,(1-4\,\ep)}
\right.\\&\left.\qquad
          - \xb^2\,\frac{9 - 12\,\ep - 101\,\ep^2 + 132\,\ep^3 + 56\,\ep^4}{18\,\ep^3\,(1-\ep)\,(1-2\,\ep)\,(1-4\,\ep)}
           \Rx
\\&
       - \Lx M_{2}(-1,-1)\,x^{2\,\ep} -  M_{3}(-1,-1)\,x^{\ep}\Rx\,\xb^{-2\,\ep}\Lx
            \frac{1-\ep - \ep^2}{9\,\ep^3\,(1-\ep)}
          + \xb\,\frac{2 + 2\,\ep}{9\,\ep^2\,(1-\ep)}
          + \xb^2\,\frac{1-\ep - \ep^2}{9\,\ep^3\,(1-\ep)}
           \Rx
\\&\left.
       + M_{5}(-1)\,x^{\ep}\,\xb^{-2\,\ep}\Lx
            \frac{1}{\ep^3}
          + \xb\,\frac{2}{\ep^2\,(1-4\,\ep)}
          + \xb^2\,\frac{1-2\,\ep-\ep^2}{\ep^3\,(1-\ep)\,(1-4\,\ep)}
           \Rx\RB
\\  +& \frac{\Gamma^3(1-\ep)\,\Gamma(1+\ep)}{\Gamma(1-2\,\ep)}\LB
         M_{0}(-1,-2)\,x^{2\,\ep}\,\xb^{-3\,\ep}\Lx
            \frac{30 - 293\,\ep + 1610\,\ep^2 - 4587\,\ep^3 + 5746\,\ep^4 - 1794\,\ep^5 - 280\,\ep^6}{18\,\ep^3\,(1-\ep)^2\,(1-2\,\ep)\,(1-4\,\ep)\,(3-2\,\ep)}
\right.\right.\\&
          + \xb\,\frac{132 - 1378\,\ep + 5204\,\ep^2 - 7950\,\ep^3 + 3782\,\ep^4 + 530\,\ep^5 - 536\,\ep^6}{9\,\ep^2\,(1-\ep)^2\,(1-2\,\ep)\,(1-3\,\ep)\,(1-4\,\ep)\,(3-2\,\ep)}
\\&\qquad\left.
          + \xb^2\,\frac{60 - 556\,\ep + 1367\,\ep^2 - 1085\,\ep^3 - 34\,\ep^4 + 202\,\ep^5}{18\,\ep^3\,(1-2\,\ep)\,(1-3\,\ep)\,(2-3\,\ep)\,(3-2\,\ep)}
           \Rx
\\&
       + M_{0}(-1,-2)\,x^{2\,\ep}\,\xb^{-3\,\ep}\,N_{f}\,\Lx
            \frac{(1-\ep)}{3\,\ep^2\,(1-2\,\ep)\,(3-2\,\ep)}
          + \xb\,\frac{2 - 2\,\ep}{3\,\ep\,(1-2\,\ep)\,(1-3\,\ep)\,(3-2\,\ep)}
\right.\\&\left.\qquad
          + \xb^2\,\frac{2 - 5\,\ep + 4\,\ep^2 - \ep^3}{3\,\ep^2\,(1-2\,\ep)\,(1-3\,\ep)\,(2-3\,\ep)\,(3-2\,\ep)}
           \Rx
\\&
       + M_{1}(-1,-2)\,x^{2\,\ep}\,\xb^{-3\,\ep}\Lx
            \frac{9 - 77\,\ep + 117\,\ep^2 + 216\,\ep^3 - 421\,\ep^4 - 60\,\ep^5}{9\,\ep^3\,(1-\ep)^2\,(1-2\,\ep)\,(1-4\,\ep)}
          - \xb\,\frac{16 - 132\,\ep + 268\,\ep^2 + 88\,\ep^3}{9\,\ep^2\,(1-\ep)\,(1-2\,\ep)\,(1-4\,\ep)}
\right.\\&\left.\qquad
          + \xb^2\,\frac{9 - 24\,\ep - 37\,\ep^2 + 75\,\ep^3 + 28\,\ep^4}{9\,\ep^3\,(1-\ep)\,(1-2\,\ep)\,(1-4\,\ep)}
           \Rx
\\&
       + M_{4}(-1,-1,-2)\,x^{2\,\ep}\,\xb^{-3\,\ep}\Lx
          - \frac{1-\ep - \ep^2}{9\,\ep^3\,(1-\ep)}
          - \xb\,\frac{2 + 2\,\ep}{9\,\ep^2\,(1-\ep)}
          - \xb^2\,\frac{1-\ep - \ep^2}{9\,\ep^3\,(1-\ep)}
           \Rx
\\&\left.\left.
       + M_{4}(1,-1,-2)\,x^{2\,\ep}\,\xb^{-3\,\ep}\Lx
          - \frac{2}{\ep^3}
          - \xb\,\frac{4}{\ep^2\,(1-4\,\ep)}
          - \xb^2\,\frac{2 - 4\,\ep - 2\,\ep^2}{\ep^3\,(1-\ep)\,(1-4\,\ep)}
           \Rx\RB\right\}
\end{split}
\end{equation}
}

\subsubsection{Contributions starting at \nnnlo}
The contributions from the square of the one-loop amplitudes starts at
\nnnlo.  The full result is too lengthy to report here, but is given,
along with assorted moments in $\xb\!$, in the supplemental material
attached to this article.  I present below only the soft contributions
(that is, the $\delta$ function and plus-distribution terms).

{\small
\begin{equation}
\begin{split}
  \sigma_{gg\to H\,g}^{3,B,{\rm soft}} =\ &
     \Cfact\,C_A^3\,\abarex^3\mumh^{3\,\ep}\left\{ \vphantom{\frac{9}{\ep^2}}
       \frac{1}{\ep^6}\,
         \frac{23}{72}\,\delta(\xb\!)
     + \frac{1}{\ep^5}\,\LB
         \frac{23}{72}\,\delta(\xb\!)
       - \frac{19}{24}{\cal D}_{0}(\xb\!)\,
        \RB
\right.\\+&
       \frac{1}{\ep^4}\,\LB
         \delta(\xb\!)\,\Lx
            \frac{23}{72}
          - \frac{247}{144}\,\zeta(2)
          \Rx
       - \frac{19}{24}\,{\cal D}_{0}(\xb\!)
       + \frac{9}{4}\,{\cal D}_{1}(\xb\!)
        \RB
     + \frac{1}{\ep^3}\,\LB
         \delta(\xb\!)\,\Lx
            \frac{127}{144}
          - \frac{247}{144}\,\zeta(2)
          - \frac{125}{36}\,\zeta(3)
          \Rx
\right.\\&\left.
       + {\cal D}_{0}(\xb\!)\,\Lx
          - \frac{19}{24}
          + \frac{275}{48}\,\zeta(2)
          \Rx
       + \frac{9}{4}\,{\cal D}_{1}(\xb\!)
       - \frac{15}{4}\,{\cal D}_{2}(\xb\!)
        \RB
\\+&
       \frac{1}{\ep^2}\,\LB
         \delta(\xb\!)\,\Lx
            \frac{185}{72}
          - \frac{247}{144}\,\zeta(2)
          - \frac{125}{36}\,\zeta(3)
          + \frac{3029}{384}\,\zeta(4)
          \Rx
       + {\cal D}_{0}(\xb\!)\,\Lx
          - \frac{49}{24}
          + \frac{275}{48}\,\zeta(2)
          + \frac{269}{24}\,\zeta(3)
          \Rx
\right.\\&\left.
       + {\cal D}_{1}(\xb\!)\,\Lx
            \frac{9}{4}
          - \frac{169}{8}\,\zeta(2)
          \Rx
       - \frac{15}{4}\,{\cal D}_{2}(\xb\!)
       + \frac{29}{6}\,{\cal D}_{3}(\xb\!)
        \RB
\\+&
       \frac{1}{\ep}\,\LB
         \delta(\xb\!)\,\Lx
            \frac{937}{144}
          - \frac{1151}{288}\,\zeta(2)
          - \frac{125}{36}\,\zeta(3)
          + \frac{3029}{384}\,\zeta(4)
          - \frac{553}{20}\,\zeta(5)
          + \frac{2125}{72}\,\zeta(2)\,\zeta(3)
          \Rx
\right.\\&
       + {\cal D}_{0}(\xb\!)\,\Lx
          - \frac{139}{24}
          + \frac{275}{48}\,\zeta(2)
          + \frac{269}{24}\,\zeta(3)
          - \frac{3841}{128}\,\zeta(4)
          \Rx
       + {\cal D}_{1}(\xb\!)\,\Lx
            \frac{21}{4}
          - \frac{169}{8}\,\zeta(2)
          - \frac{171}{4}\,\zeta(3)
          \Rx
\\&\left.
       + {\cal D}_{2}(\xb\!)\,\Lx
          - \frac{15}{4}
          + \frac{335}{8}\,\zeta(2)
          \Rx
       + \frac{29}{6}\,{\cal D}_{3}(\xb\!)
       - \frac{21}{4}\,{\cal D}_{4}(\xb\!)
        \RB
\\+&
         \delta(\xb\!)\,\Lx
            \frac{547}{36}
          - \frac{1561}{144}\,\zeta(2)
          - \frac{1193}{144}\,\zeta(3)
          + \frac{3029}{384}\,\zeta(4)
          - \frac{553}{20}\,\zeta(5)
          + \frac{2125}{72}\,\zeta(2)\,\zeta(3)
\right.\\&\left.
          - \frac{84281}{3072}\,\zeta(6)
          + \frac{4607}{144}\,\zeta(3)^2
          \Rx
       + {\cal D}_{0}(\xb\!)\,\Lx
          - \frac{349}{24}
          + \frac{593}{48}\,\zeta(2)
          + \frac{269}{24}\,\zeta(3)
          - \frac{3841}{128}\,\zeta(4)
\right.\\&\left.
          + \frac{4869}{40}\,\zeta(5)
          - \frac{5581}{48}\,\zeta(2)\,\zeta(3)
          \Rx
       + {\cal D}_{1}(\xb\!)\,\Lx
            \frac{57}{4}
          - \frac{169}{8}\,\zeta(2)
          - \frac{171}{4}\,\zeta(3)
          + \frac{6777}{64}\,\zeta(4)
          \Rx
\\&
       + {\cal D}_{2}(\xb\!)\,\Lx
          - \frac{31}{4}
          + \frac{335}{8}\,\zeta(2)
          + \frac{373}{4}\,\zeta(3)
          \Rx
       + {\cal D}_{3}(\xb\!)\,\Lx
            \frac{29}{6}
          - \frac{701}{12}\,\zeta(2)
          \Rx
       - \frac{21}{4}\,{\cal D}_{4}(\xb\!)
       + \frac{149}{30}\,{\cal D}_{5}(\xb\!)
\\&\left.
     + \cal{O}(\ep)\vphantom{\frac{225}{\ep^6}}\right\}
\\  \sigma_{q\overline{q}\to H\,g}^{3,B,{\rm soft}} =\ & 0
\\  \sigma_{q\overline{q}\to H\,g}^{3,B,{\rm soft}} =\ & 0
\end{split}
\end{equation}
\section{Conclusions and Outlook}
\label{sec:conclude}
I have computed the contributions of one-loop single-real-emission
amplitudes to inclusive Higgs boson production at \nnnlo.  Though a
substantial calculation, this is but a portion of the full \nnnlo\
result.  I have computed this contribution to the cross section as an
extended threshold expansion, obtaining enough terms to invert the
series and determine the closed functional form through order $\ep^1$.
I have also computed the contributions of these same amplitudes to the
\nlo\ and \nnlo\ inclusive cross sections in closed form, in terms of
$\Gamma$-functions and the hypergeometric functions ${}_{2}F_{1}$ and
${}_{3}F_{2}$.  These functions can be readily expanded to all orders
in $\ep$.

The methods used in this calculation can be immediately applied to
other single-inclusive production processes like Drell-Yan or
pseudoscalar production.  In the current calculation, I have only
considered single-real emission contributions.  However, the basic
method was already used more than ten years ago to compute double-real
emission contributions at
\nnlo~\cite{Harlander:2002wh,Harlander:2002vv,Harlander:2003ai}.  The
phase space for triple-real emission is far more complicated than that
for single- or double-real emission and it may be that the methods of
Ref.~\cite{Anastasiou:2013srw}, working on the other side of the
Cutkosky relations and threshold-expanding cut loop integrals rather
than phase space integrals, is more effective for that process.

\vskip20pt

\paragraph*{Acknowledgments:}
I would like to thank Lance Dixon for a stimulating discussion.  I
would also like to thank the authors of Ref.~\cite{Anastasiou:2013mca}
for their assistance in comparing results.  This research was
supported by the U.S.~Department of Energy under Contract
No.~DE-AC02-98CH10886.

\appendix
\section{The Computation of the Scale-Free Integrals}
I call Master integrals $M_{0}$, $M_{2}$, $M_{10}$ and $M_{11}$
scale-free, since they do not depend on the threshold parameter
$\xb\!$ and integrate to pure numbers.  $M_{0}$, $M_{2}$ can be
integrated in closed form.  Expressions for their expansion in $\ep$
are given below.  One can obtain closed-form expressions for Integrals
$M_{10}$ and $M_{11}$, just as one could for $M_{26}$ in
\eqn{eqn:mi26}, but such expressions are difficult to expand in $\ep$.
However, these integrals can be readily computed to arbitrary order in
$\ep$ by making use of hypergeometric identities and the algebraic
properties of harmonic polylogarithms.  First, the hypergeometric
identities:
\begin{equation}
\begin{split}
\label{eqn:hypgeos}
  \hypgeo{1}{-\ep}{1-\ep}{-\frac{y}{\yb}} &=
     \yb^{-\ep}\,\hypgeo{-\ep}{-\ep}{1-\ep}{y}\,,\\
  \hypgeo{1}{-\ep}{1-\ep}{-\frac{\yb}{y}} &=
     y^{-\ep}\,\hypgeo{-\ep}{-\ep}{1-\ep}{\yb}\\
    &= 1 + y^{-\ep}\,\yb^{\ep}\,\Gamma(1-\ep)\Gamma(1+\ep)
     - \yb^{\ep}\hypgeo{\ep}{\ep}{1+\ep}{y}\,.\\
\end{split}
\end{equation}
I next expand all of these terms in powers of $\ep$, harmonic
polylogarithms of argument $y$ and $\zeta$-functions.
\begin{equation}
\begin{split}
\hypgeo{\ep}{\ep}{1+\ep}{y} &= 1 + \sum_{n=2}^{\infty}\
   \Lx-\ep\Rx^{n}\sum_{m=1}^{n-1}\ (-1)^{m-1}\,H\Lx\vec{\bf0}_{n-m},
    \vec{\bf1}_{m};y\Rx\,,\\
y^{\alpha\,\ep} &= \sum_{n=0}^\infty\Lx\alpha\,\ep\Rx^n\,
     H\Lx\vec{\bf0}_{n};y\Rx\,,\\
\yb^{\beta\,\ep} &= \sum_{n=0}^\infty\Lx-\beta\,\ep\Rx^n\,
     H\Lx\vec{\bf1}_{n};y\Rx\,.\\
\Gamma(1-\ep)\,\Gamma(1+\ep) &= \frac{\ep\,\pi}{\sin\,\ep\,\pi}
   = 1 + \sum_{n=1}^{\infty}\,\Lx2-2^{2-2\,n}\Rx\,\ep^{2\,n}\,\zeta(2\,n)\,.
\end{split}
\end{equation}
where $\vec{\bf0}_{n}$ and $\vec{\bf1}_{m}$ represent strings of $n$
$0$'s and $m$ $1$'s, respectively.  The resulting products of HPLs
can be combined into a sum of single HPLs by using the shuffle
identity as in \eqn{eqn:shuffle}.  The result is that each term
consists of a factor of $y^{-1} = f_0(y)$ multiplying a single HPL
with weight vector containing only $0$'s and $1$'s.  Finally, I use
the definition of the HPLs, \eqn{eqn:HPLint}, to obtain
\begin{equation}
  \int_0^1\ f_{0}(y)\ H(\vec{w};y) = H(0,\vec{w};1) = \zeta(0,\vec{w})\,.
\end{equation}
The result for the master integrals is
\begin{equation}
\begin{split}
M_{10} =&
         1
       - \ep^{2} \, (
            3\,\zeta(2)
          )
       - \ep^{3} \, (
            14\,\zeta(3)
          )
       - \ep^{4} \, (
            \frac{173}{4}\,\zeta(4)
          )
       - \ep^{5} \, (
            152\,\zeta(5)
          - 14\,\zeta(2)\,\zeta(3)
          )\\&
       - \ep^{6} \, (
            \frac{18083}{48}\,\zeta(6)
          - 8\,\zeta(3)^{2}
          )
       - \ep^{7} \, (
            1261\,\zeta(7)
          + \frac{117}{2}\,\zeta(3)\,\zeta(4)
          - 152\,\zeta(2)\,\zeta(5)
          )
         + {\cal O}(\ep^8)\,,\\
M_{11} =&
         1
       - \ep^{2} \, (
            3\,\zeta(2)
          )
       - \ep^{3} \, (
            14\,\zeta(3)
          )
       - \ep^{4} \, (
            \frac{157}{4}\,\zeta(4)
          )
       - \ep^{5} \, (
            126\,\zeta(5)
          - 18\,\zeta(2)\,\zeta(3)
          )\\&
       - \ep^{6} \, (
            \frac{3737}{16}\,\zeta(6)
          - 26\,\zeta(3)^{2}
          )
       - \ep^{7} \, (
            774\,\zeta(7)
          - \frac{211}{2}\,\zeta(3)\,\zeta(4)
          - 138\,\zeta(2)\,\zeta(5)
          )
         + {\cal O}(\ep^8)\,.
\end{split}
\end{equation}

\subsection{Master Integrals $M_{0}$ and $M_{2}$}
Master integrals $M_{0}$ and $M_{2}$ are known in closed form and can
be readily expanded in $\ep$.  
\begin{equation}
\label{eqn:M0exp}
M_{0}(\alpha,\beta) =
    \frac{\Gamma(1+\alpha\,\ep)\,\Gamma(1+\beta\,\ep)}{\Gamma(1+(\alpha+\beta)\,\ep)}
   = \exp\LB-\sum_{n=2}^\infty\ \,\Lx\sum_{m=1}^{n-1}\binom{n}{m}
        \alpha^m\,\beta^{n-m}\Rx\frac{(-\ep)^{n}\,\zeta(n)}{n}\RB
\end{equation}
\begin{equation}
\begin{split}
M_{2}&(\alpha,\beta) = M_{0}(\alpha,\beta-1)
     \ghypgeo{-\ep}{-\ep}{\alpha\,\ep}{1-\ep}{1+(\alpha+\beta-1)\ep}{1}\\
    =&\ M_{0}(\alpha,\beta-1)\left\{1
       + \ep^{3}\,\zeta(3) \,\alpha
       + \ep^{4}\,\zeta(4) \,\Lx
            2\,\alpha
          - \frac{5}{4}\,\alpha\,\beta
          - \alpha^2
          \Rx\right.
\\&
       + \ep^{5}\LB\zeta(5) \,\Lx
            3\,\alpha
          - \frac{3}{2}\,\alpha\,\beta
          - \frac{1}{2}\,\alpha\,\beta^2
          + \frac{5}{2}\,\alpha^2
          - \frac{3}{2}\,\alpha^2\,\beta
          + \alpha^3
          \Rx
\right.\\&\left.\qquad
       + \zeta(3)\,\zeta(2) \,\Lx
          - \alpha\,\beta
          + \alpha\,\beta^2
          - 3\,\alpha^2
          + 2\,\alpha^2\,\beta
          \Rx\vphantom{\frac{3}{2}}\RB
\\&
       + \ep^{6}\LB\zeta(6) \,\Lx
            4\,\alpha
          - \frac{61}{12}\,\alpha\,\beta
          + \frac{101}{48}\,\alpha\,\beta^2
          - \frac{1}{12}\,\alpha\,\beta^3
          - \frac{17}{6}\,\alpha^2
          + \frac{67}{48}\,\alpha^2\,\beta
          - \frac{1}{4}\,\alpha^2\,\beta^2
          + \frac{13}{6}\,\alpha^3
          - \frac{23}{12}\,\alpha^3\,\beta
          - \alpha^4
          \Rx
\right.\\&\left.\qquad
       + \zeta(3)^2 \,\Lx
          - \alpha\,\beta
          + 2\,\alpha\,\beta^2
          - \alpha\,\beta^3
          - \frac{5}{2}\,\alpha^2
          + \frac{11}{2}\,\alpha^2\,\beta
          - \frac{5}{2}\,\alpha^2\,\beta^2
          + \frac{3}{2}\,\alpha^3
          - \alpha^3\,\beta
          \Rx\RB
\\&
       + \ep^{7}\LB\zeta(7) \,\Lx
            5\,\alpha
          - 5\,\alpha\,\beta
          - \frac{19}{16}\,\alpha\,\beta^2
          + \frac{99}{16}\,\alpha\,\beta^3
          - 4\,\alpha\,\beta^4
          + 7\,\alpha^2
          - \frac{115}{8}\,\alpha^2\,\beta
          + \frac{179}{8}\,\alpha^2\,\beta^2
\right.\right.\\&\left.\hskip 60pt
          - 13\,\alpha^2\,\beta^3
          + \frac{29}{16}\,\alpha^3
          + \frac{211}{16}\,\alpha^3\,\beta
          - 12\,\alpha^3\,\beta^2
          + 7\,\alpha^4
          - 5\,\alpha^4\,\beta
          + \alpha^5
          \Rx
\\&\qquad
       + \zeta(5)\,\zeta(2) \,\Lx
          - \alpha\,\beta
          + 2\,\alpha\,\beta^2
          - 2\,\alpha\,\beta^3
          + \alpha\,\beta^4
          - 5\,\alpha^2
          + \frac{11}{2}\,\alpha^2\,\beta
          - \frac{11}{2}\,\alpha^2\,\beta^2
          + 4\,\alpha^2\,\beta^3
\right.\\&\left.\hskip 80pt
          - \frac{5}{2}\,\alpha^3
          - \frac{9}{2}\,\alpha^3\,\beta
          + 6\,\alpha^3\,\beta^2
          - 5\,\alpha^4
          + 4\,\alpha^4\,\beta
          \Rx
\\&\qquad
       + \zeta(3)\,\zeta(4) \,\Lx
          - 3\,\alpha\,\beta
          + \frac{29}{4}\,\alpha\,\beta^2
          - \frac{29}{4}\,\alpha\,\beta^3
          + 3\,\alpha\,\beta^4
          - 7\,\alpha^2
          + \frac{85}{4}\,\alpha^2\,\beta
          - \frac{97}{4}\,\alpha^2\,\beta^2
\right.\\&\left.\left.\hskip 80pt
          + 9\,\alpha^2\,\beta^3
          + \frac{41}{4}\,\alpha^3
          - \frac{35}{2}\,\alpha^3\,\beta
          + 7\,\alpha^3\,\beta^2
          - 3\,\alpha^4
          + 2\,\alpha^4\,\beta
          \Rx\RB
\\&
        + {\cal O}(\ep^8)\left.\vphantom{\frac{3}{4}}\right\}\,.\\
\end{split}
\end{equation}

\end{document}